%% file: main.tex
\documentclass{svproc}
\usepackage{graphicx}%
\usepackage{multirow}%
\usepackage{amsmath,amssymb,amsfonts}%
\usepackage{mathrsfs}%
\usepackage[title]{appendix}%
\usepackage{xcolor}%
\usepackage{textcomp}%
\usepackage{manyfoot}%
\usepackage{booktabs}%
\usepackage{algorithm}%
\usepackage{algorithmicx}%
\usepackage{algpseudocode}%
\usepackage{listings}%
\usepackage{hyperref}

\usepackage{cite}
\usepackage{dsfont}
\usepackage{siunitx}
\usepackage{multirow}
\usepackage{makecell,multirow}
\usepackage{array}
\usepackage{subcaption}
\usepackage[margin=1in]{geometry}

\newcommand{\ie}{{\em i.e.,} }
\newcommand{\smallast}{{\scriptsize *}}
\newcommand{\smallastdub}{{\scriptsize **}}
\newcommand{\smallasttrip}{{\scriptsize ***}}

\usepackage{url}

\begin{document}
\mainmatter              % start of a contribution
%
%\title{Quantile Autoregressive Networks for estimating probabilistic Causal Variation in Cryptocurrency Markets}
%
%\title{Quantile Vector Autoregression for estimating distributional causality networks. }

\title{Piecewise Linear Quantile Vector Autoregression for Distributional Causality Networks}

%matts suggestio
%\title{Piecewise Linear Vector Autoregression for inferring Networks}

\title{Enhancing Causal Discovery in Financial Networks with Piecewise Quantile Regression}

%\title{Quantile Autoregressive Networks for estimating probabilistic Causal Variation in Cryptocurrency Markets} --- "full spectrum"?

%\titlerunning{Piecewise Quantile Vector Autoregression}  % abbreviated title (for running head)
% \author{Author1Name Author1Surname\inst{1} \and Author2Name Author2Surname\inst{2}
% Author3Name Author3Surname \and Author4Name Author4Surname \and Author5Name Author5Surname \and Author6Name Author6Surname}
\author{Cameron Cornell\inst{1}, Lewis Mitchell\inst{1},
Matthew Roughan\inst{1}}

\authorrunning{Cornell, Mitchell, Roughan} % abbreviated author list (for running head)
%

%\institute{Institution1, address,\\
%\email{author1name.author1surname@institution.email},\\ WWW home page:
%\texttt{http://users.com/author1.html}
%\and
%Institution2,
%address,\\
%state, country}
\institute{The University of Adelaide, Adelaide, Australia,\\
\email{\{cameron.cornell,lewis.mitchell,matthew.roughan\}@adelaide.edu.au}\\ 
}

\maketitle              % typeset the title of the contribution

\begin{abstract}
Financial networks can be constructed using statistical dependencies found within the price series of speculative assets. %Across the various methods used to infer these networks, there is a general reliance on predictive modelling to capture cross-correlation effects, usually seeking to model the flow of mean-response information, or to reveal the propagation of volatility and risk within the market.
Across the various methods used to infer these networks, there is a general reliance on predictive modelling to capture cross-correlation effects. These methods usually %seek to
model the flow of mean-response information, or %to reveal
the propagation of volatility and risk within the market. Such techniques, though insightful, don't fully capture the broader distribution-level causality that is possible within speculative markets.  This paper introduces a novel approach, combining quantile regression with a piecewise linear embedding scheme --- allowing us to construct causality networks that identify the complex tail interactions inherent to financial markets. Applying this method to 260 cryptocurrency return series, we uncover significant tail-tail causal effects and substantial causal asymmetry. We identify %high `self-causality'
a propensity for coins to be self-influencing, with comparatively sparse cross variable effects. Assessing all link types in conjunction, Bitcoin stands out as the primary influencer --– a nuance that is missed in conventional linear mean-response analyses. Our findings introduce a comprehensive framework for modelling distributional causality, paving the way towards more holistic representations of causality in financial markets. %\keywords{keyword1, keyword2, keyword3}
\end{abstract}
\section{Introduction}
Understanding the dynamics that lead to price fluctuations in speculative %assets holds significance
is critical for a variety of decision-makers.  It allows both investors and active market participants (traders) to make more informed transactions --- further providing both participants and policymakers with a deeper comprehension of the risks associated with market participation. A crucial aspect of market dynamics is the concept of price causality, which delves into the temporal dependencies between assets. %However, a complete picture of this causality is difficult to achieve, and current methodologies provide only a narrow glimpse into the true dynamics. 
%provides limited insight into the true dynamics.
However, achieving a complete understanding of these causal relationships remains challenging, as existing methodologies often provide only a narrow view of the true interactions. In this paper, we develop probabilistic forecasting methodologies to infer a network representation of the causal dynamics of a system, and we apply our methodology to a cryptocurrency dataset. 
%dynamics of cryptocurrency markets.

% is this even meant to remain in the paper??
%This insight is invaluable to investors and %active market participants (
%traders %)
%as it empowers them to execute more informed transactions. Additionally, both individuals and policymakers benefit from a deeper comprehension of the risks tied to market participation.
%It can assist investors or active market participants (traders) to make more informed transactions. Further, both these individuals and policymaker benefit from a deeper comprehension of the risks tied to market participation. 

The recent emergence of cryptocurrency markets provides a fertile environment for modelling large-scale causal interactions. These assets rapidly rose to public prominence following their remarkable growth %in the preceding decade
. This growth, however, is often viewed as highly unstable, with prices showing extreme levels of volatility and responsivity to social and economic factors --- such as market sentiment, news events, technological advancements, regulatory changes, and the erratic behaviour of market participants. Many speculate that these emerging markets are less efficient than traditional, regulated financial markets --- a view supported by several studies \cite{ALYAHYAEE2018228}. While the consequences of such inefficiency remain unclear, it stands to incentivize the use of causal analysis.

A common way to investigate causal dynamics in financial markets involves using Vector Autoregression (VAR) to model linear mean-response effects \cite{cornell2023vector, rank2, conf}. The method is a multivariate expansion of the standard univariate autoregression to account for cross-dependencies, allowing us to construct networks that indicate cross-causal effects. Another frequent approach is to use Generalised Autoregressive Conditional Heteroskedastic (GARCH) techniques to model the causal relationships in return volatility, which can similarly be cast into a network structure \cite{rank2,conf}. This approach can be extended to higher-order moments (such as skew), with the combined results providing insight into causal dynamics across the target distribution. 

%with the combined results providing insight into the overall causal dynamics. 

%This approach can be extended to encompass higher-order moments as well, and when combined, these results offer a more comprehensive insight into the overall distributional causality structures within financial markets.

%For the purposes of distributional representation, these Linear - Moment models have several inherent limitations. Firstly, the modelling of separate moments (mean, volatility, skew and kurtsosis etc) independently makes it difficult to discern net variable effects. Secondly, the modelling of higher order moments is generally challenged due to compounding outlier effects. Lastly, the linear, parametric form of these techniques affords low flexibility to complex relationships. 

However, in the context of distributional representations, these linear-moment models exhibit several inherent limitations. Modelling the moments (mean, volatility, skew, and kurtosis, etc.) independently hinders the discernment of net variable effects, and higher order moments compound outlier effects, leading to estimation difficulties. These limitations are most pronounced in the distributional tails, which are of great importance in volatile financial markets, and tend to be the key region of interest. %and are often the region of interest in academic studies.  

%Beyond these identification issues, the linear, parametric form of these techniques
%has little flexibility in capturing complex relationships. 

%matts comments
%however, tial behaviuor is of great importance in volatile financial markets, and so approaches that consider distributional representations have advantages. coupled with this is the problem introduced by outlier and estimates of standard models?

%Integrating both forms of analysis, 

%We propose an alternative, quantile regression based approach which captures mean-response, volatility and higher-moment causality in a single model, %--- further highlighting 
%allowing understanding of asymmetries and tail-event interactions. The emphasis on these effects, bolstered by the robust estimates produced from probabilistic prediction renders our model particularly suitable for representing financial market data. 

%We propose an alternative, quantile regression based approach which captures mean-response, volatility and higher-moment causality in a single model. %--- further highlighting 
%Our model, titled Piecewise Quantile Vector Autoregression (P-QVAR) 

%allowing understanding of asymmetries and tail-event interactions. The emphasis on these effects, bolstered by the robust estimates produced from probabilistic prediction renders our model particularly suitable for representing financial market data. 

Our proposed method, Piecewise Quantile Vector Autoregression (P-QVAR) captures mean-response, volatility and higher-moment causality in a single model, revealing asymmetries and tail-event interactions. The emphasis on these effects, bolstered by the robust estimates produced from probabilistic prediction renders our model particularly suitable for representing financial market data.

The primary contributions of this paper include:
\begin{itemize}
    \item[\textbullet] The introduction of a novel probabilistic model that combines quantile regression with piecewise linear embeddings to construct causal networks capable of capturing complex tail-tail interactions.
    
    \item[\textbullet] The empirical application of P-QVAR to one year of hourly return data from 260 cryptocurrencies, yielding distributional causality networks that further our understanding of the complex, asymmetric causal flow inherent to cryptocurrency markets. 

    \begin{itemize}
        \item[-] Our results indicate that the majority of causal links involve the tail regions, with 73\% originating from tails, 84\% targeting the tails, and 60\% being tail-tail interactions. 
        
        %\item[-] We observe that the majority of causal links are self-influencing, with only xyz\% of links indicating the cross-causality of coin pairs.

        \item[-] While the set of self-causal links are more densely depopulated than their cross-causal counterparts (65.8\% vs 4.5\%), the vast majority (94.7\%) of statistically significant links are cross-causal. 
        
        %(65.8\% vs 4.5\% for cross causal)

        %\item[-] We demonstrate that non-distributional causality networks fails to identify xyz\% of the links our method observes.  

        \item[-] We demonstrate that P-QVAR identifies 6.5 times as many effects as standard VAR. 
        %fails to identify xyz\% of the links our method observes.  
    \end{itemize}

    %\item[\textbullet] The introduction of a visualisation tool: 'Quantile Influence Graph' for discretely representing distributional causality.

    %Development of a new visualization tool, the 'Quantile Influence Graph', designed for discrete representation of distributional causality

    \item[\textbullet] The introduction of a new visualisation tool, the `Quantile Influence Graph', designed %for discrete representation of distributional causality.
    as a graphical representation of distributional causality. 
\end{itemize}

\section{Related Work}

%financial network
Networks based on correlational structures in financial asset time series data have been explored in many papers. Often, researchers investigate simultaneous correlations between prices, seeking to model the joint structure of these observations as a network \cite{stat_analysis_of_network}. Several studies analyze the temporal evolution of these networks using sliding-window  methods \cite{partial_corr,structural_entropy}, with others investigating evolution by incrementally adding nodes based on their correlations, developing an asset graph \cite{clustering_and_info}.  

%causal/partial effects
Causal networks can be generated from cross-correlational effects. These types of networks have been constructed before \cite{billio2011} and are frequently used to model the joint causal dynamics of price and sentiment \cite{connecting_emotions,Combiningsocial_and_financial}. Cross-correlational effects are typically modelled in a partial-effects framework to control for confounding variables. The VAR model is a common approach for this purpose, and is equivalent to partial effects of the cross-correlation matrix. Variations to this model have been developed to incorporate specific features, such as long range dependency 
\cite{long_range_VAR} and restrictions to acyclic graphs \cite{ahelegebey}. 

%cryptocurrency specific modelling
Cryptocurrency causality network research has generally focused on the interactions between cryptocurrencies and other data (sentiment, traditional financial assets, {\em etc.}) \cite{connecting_emotions,crypto1, amirzadeh_nazari_thiruvady_ee, Elsayed2020CausalityAD, AZQUETAGAVALDON2020122574, Ahelegbey2021NetworkBE}. Beyond a small number of studies ~\cite{Ahelegbey2021NetworkBE, cornell2023vector} the majority of research in this space employs bivariate analysis, rather than the full partial-effects found in VAR models. 

%GARCH
%Compared to standard VAR, there is comparatively dense studies on the modelling of cryptocurrency markets with GARCH models (a loose VAR equivalent for volatility), however these are generally not framed in terms of network analysis, and usually involve a limited number of variables. 

Compared to standard VAR, there is a more substantial body of research on modeling cryptocurrency markets with GARCH models (which can be considered a loose VAR equivalent for volatility). However, these studies are generally not framed in terms of network analysis, and usually involve a limited number of variables. A common theme in these studies is the identification of assymetric effects \cite{chu, CHEIKH2020101293, FAKHFEKH2020101075, GARCHN}, such as treating positive and negative returns separately.

Quantile regression has been employed to model the autoregressive components of economic data across different target quantiles \cite{QAR}. Several econometric studies have analyzed `quantile-causality' by assessing covariate effects across a dense set of targets, fitting 100 regressions and observing how coefficients vary across the target range \cite{BAUR2013786,CHUANG20091351}, with the latter incorporating some nonlinearity by including both raw and squared return covariates. However, this type of dense analysis is not be suitable for our objectives due to its poor scaling in relation to our desired network size. Other studies focus on a select set of representative quantiles, often emphasizing asymmetric effects by separating variable effects into positive and negative regions \cite{assymetric1, assymetric2}. 
%Extreme value asymmetries in the context of quantile-causality have been explored by fitting independent linear regions \cite{covid_indicator, GUO2018251} --- a specification that slightly deviates from the contiguous construction traditionally linked with piecewise linear regression, which we employ in this paper. 
Extreme value asymmetries in the context of quantile-causality have been explored by fitting independent linear regions \cite{covid_indicator, GUO2018251}. This specification deviates from the contiguous construction traditionally linked with piecewise linear regression, which we employ in this paper. More generalized nonlinear causal   effects for quantile causality have been explored using kernel-based estimation techniques, with several applications to financial data \cite{BALCILAR201674, JENA2019615}.

While quantile regression's application to network modelling has been somewhat limited, one significant study extended the recently developed ``Network Autoregression" \cite{NAR} into a quantile regression framework \cite{ZHU2019345}. However, these approaches employ spatial modelling techniques, without a heavy emphasis on causality. This study bridges the gap between the recent advances in causality analysis with complex network based representations. We introduce a model that can identify the type of complex causal patterns that have seen recent academic focus, while also retaining a high degree of interpretability, allowing for a direct representation in a network structure. %Our model produces highly visualisable relationships, which we demonstrate in our empirical study. 
Our model allows for highly intuitive visual representations, which we demonstrate in our empirical study. 

%We demonstrate the ease of interpretibility with several visualiations that 

%reference the markov switching VAR

%%% compile some citations first 

%\cite{BAUR2013786} has continious target, with linear covariate effect. co-efficient is based on the average of the per quantile effects. some weird binary indicator idk. 

%\cite{CHUANG20091351} has continious targets, once again multitesting the set of beta co-efficients. Takes the sqaured returns into account for one of the models. 

%\newpage
%\section{Methodology/Constructing Networks}
\section{Network Construction Methdologoy}
\noindent The aim of causal network analysis is to construct a directed graph (digraph) denoted by $G=(V,E)$, where $V$ is a set of nodes and $E$ is a set of edges representing causal dependencies. Each edge $e_{i,j} \in E$ indicates that the next observation of asset $j$ depends on the previous values of $i$. The set of edges $E$ may be represented as an $N \times N$ adjacency matrix $W$ (where $N$=$|V|$ is the number of price series under study), with elements $W_{i,j}=1$ if there is a link $e_{i,j}$, and $W_{i,j}=0$ otherwise. 

The primary methodological choice when developing these networks from empirical data is the selection of a suitable model to test the interrelations within the time series, \ie  to infer $E$. This section provides an overview of the common methodology for this estimation problem (VAR), as well our proposed extension, P-QVAR.  

\subsection{Vector Autoregression}
\label{sec:VAR}
Vector autoregression (VAR) is a popular statistical model introduced by the macroeconometrician Christopher Sims \cite{sims_1980} to model the joint dynamics and causal relations among a collection of time series. It is the natural multivariate extension of the univariate autoregression (AR) model frequently used to analyse the inter-temporal dependency of a sequence of observations. Under the VAR(p) formulation the expectation of the data vector $\boldsymbol{y}_t$ at the next observation is a linear function of $p$ previous observations. Equations 1 and 2 below show the relationship for order-1 and order-p lagged variants: %within a system of $N$ variables  
%\begin{align}
%\bf{y}_t &=A\bf{y}_{t-1}+c+\bf{\epsilon}_t \\
%\textbf{y}_t &=A_1\textbf{y}_{t-1}+A_2\textbf{y}_{t-2}+...+A_{t-p}\textbf{y}_{t-%p}+c+\mathbf{\epsilon}_t. 
%\end{align}
\begin{align}
\mbox{Order-1:} \;\;\; & \boldsymbol{y}_t =A_1\boldsymbol{y}_{t-1}+\boldsymbol{c}+\boldsymbol{\epsilon}_t , \label{eq:var_lin} \\
\mbox{Order-$p$:} \;\;\; & \boldsymbol{y}_t =A_1\boldsymbol{y}_{t-1}+A_2\boldsymbol{y}_{t-2}+...+A_{t-p}\boldsymbol{y}_{t-p}+\boldsymbol{c}+\boldsymbol{ \epsilon }_t , 
\end{align}

\noindent where $\boldsymbol{y}_t$ is a $N \times 1$ vector of observations at time $t$, $\boldsymbol{c}$ is a constant vector, the $A_k$ are $N \times N$ coefficient matrices for lags $k=1,...,p$, and $\boldsymbol{\epsilon}_t$ is a $N \times 1$ vector of error terms with zero mean and covariance matrix $\Sigma_\epsilon$. The vector $\boldsymbol{\epsilon}_t$  is often assumed to be Gaussian, however this is not a strict requirement. The VAR model assumes that the current value of each variable depends on its past values as well as the past values of potentially all other variables in the system (full conditioning). 

%\textcolor{myblue}{The matrices $A_i$ fully characterise the causal structure of our model.}

The estimation of a VAR model comprises estimating the coefficient matrices $A_k$ and the error covariance matrix $\Sigma_\epsilon$. As we are interested in the causal influence structure within our dataset, we primarily require estimates of $A_k$, as they fully characterise the causal relations. This is often accomplished by the multivariate least squares (MLS) approach under which estimating the VAR is viewed as a general multivariate regression problem, with closed-form solutions generated via orthogonal projection \cite{luet}.

We may conduct hypothesis tests for the statistical significance of the elements of the coefficient matrices by noting that our estimates $\hat{A_k}$ are asymptoptically normally distributed under finite variance assumptions, \ie 
\begin{align}
    \sqrt{N} \; \mbox{Vec}(\hat{A}_k-A_k)\xrightarrow[]{d}\mathcal{N}(0, \Gamma^{-1}\otimes\Sigma_\epsilon) , \label{eq:kron}
\end{align}
where $\Gamma=\ YY'/N$, $\otimes$ indicates the Kronecker product and $\mbox{Vec}(\cdot)$ denotes casting a matrix into vector form. For the case of a VAR(1) model the term $Y$ is the matrix representation of our response data $\boldsymbol{y}_t$, implying that $\Gamma$ is an estimate of the covariance matrix of returns. For generalised VAR(p) the complexity of this matrix increases, however Equation \ref{eq:kron} is still valid. 
%Hence, we may construct $t_{i,j}$ values associated with the null hypothesis $A_{k,i,j}=0$ as $\hat{A}_{k,i,j}/\hat{s}_{i,j}$, with $\hat{s}_{i,j}$ being the relevant term from $\Gamma^{-1}\otimes\Sigma_\epsilon$. 
To establish the existence of link $e_{i,j}$ we construct $t$ values associated with the null hypothesis $A_{i,j}=0$ as $t_{i,j}=\hat{A}_{i,j}/\hat{s}_{i,j}$, where $\hat{s}_{i,j}$ is the relevant term from $\Gamma^{-1}\otimes\Sigma_\epsilon$. These $t$ vales can be used to generate a binary link classification with a false positive probability $\alpha$ as $e_{i,j}=|t_{i,j}|>\Phi(1-\alpha/2)$, where $\Phi$ is the inverse normal CDF. For the empirical networks discussed in \autoref{sec:results}
, we select $t$ values corresponding to $\alpha=0.001$, providing a relatively high detection threshold.

% For the detection of edges in our empirical networks we chose $t$ such that our implied false positiv

To streamline our discussion and estimation of causal networks, we limit our analysis to VAR(1) processes and omit the index $p$ from our discussion, with $A=A_1$ unless otherwise specified. %For simulations, we note that VAR(p) processes can be transformed into a VAR(1) form \cite{luet}, implying that our VAR(1) simulation results should generalise as the fitting routines remain unchanged.
For the empirical networks in Section \ref{sec:results} we operate under the assumption that any causal link $i\rightarrow j$  will first manifest in order-1 effects, and that lag $p>1$ effects will not occur independently of a 
$p=1$ dependence. This assumption is intuitive and the scenarios where it doesn't hold are expected to be relatively rare.

\subsection{Quantile Regression}
%Quantile regression is a probabilistic model that has been used across many fields to represent linear variable effects on the full conditional density. The simplest representaiton of quantile regression is the least absolute deviation (LAD) regression, where the minimisation of the \(\mathcal{L}_1\) residual loss function, $|y_t-\hat{y_t}|$, results in an estimate of the conditional median. By further altering the loss function into the following form: 

% Quantile regression (QR) is a probabilistic model that has seen widespread application to model linear variable effects across distinct regions of the conditional target density, $F_{Y|X}(y)$. %$F_{Y_t|X_{t-1}}(y_t)$.
% The simplest, and first introduced format of this tool is the Least Absolute Deviation (LAD) regression, where minimisation of the \(\mathcal{L}_1\) residual loss function: $|y_t-\hat{y_t}|$ results in an estimate of the conditional median (50th quantile). %By further altering the \(\mathcal{L}_1\) residual loss into the following 'pinball loss' format:
% The 'Full' QR format arises when extending the \(\mathcal{L}_1\) residual loss into the following 'pinball loss' format:

Quantile regression (QR) is a probabilistic approach widely used to model linear variable effects across distinct regions of the conditional target density, $F_{Y|X}(y)$. This technique was originally developed in the context of the Least Absolute Deviation (LAD) regression, where linear models trained by minimizing the \(\mathcal{L}_1\) residual loss function: $|y_t-\hat{y}_t|$ produce estimates of the conditional median (50th quantile). Full QR, targeting some arbitrary distributional quantile $q$,  is achieved by extending the \(\mathcal{L}_1\) residual loss to the following pinball loss format:

\begin{equation}
\label{eqn:pinball_loss}
    \mathcal{L}_{q}(y_t,\hat{y}_{t,q}) = 
\begin{cases} 
      (y_t - \hat{y}_{t,q}) \times q & \text{for } \hat{y}_{t,q} \leq y_t \\
      (\hat{y}_{t,q} - y_t) \times (1-q) & \text{for } y_t < \hat{y}_{t,q}. \\
   \end{cases}
\end{equation}
A regression model trained to minimise the expected pinball loss \(\mathcal{L}_q\) for a given quantile $ q\in (0, 1) $ can be shown to result in predictions of the conditional quantile $Q_{Y|X}(q)$. %The predictions for this conditional quantile are generated as a linear combination of the $i$ predictor variables:
Specifically, these predictions are generated as a linear combination of the $j$ predictor variables:
\begin{equation}
    \hat{y}_{t,q} = \hat{Q}_{Y_t|X_{t-1}}(q) = \sum^{n}_{j}\alpha_jx_{t-1,j}+c = \boldsymbol{\hat{\alpha}}^T X_{t-1}+c,
\end{equation}
where the fitted values $\boldsymbol{\hat{\alpha}}$ are determined from the following empirical minimisation on the $T$ training observations:
\begin{equation}
    \boldsymbol{\hat{\alpha}},c= \operatorname*{argmin}_{\alpha,c} \sum^T_{t=1} \mathcal{L}_q(y_t, \boldsymbol{\hat{\alpha}}^T X_t+c). \label{eq:armin}
\end{equation}

\noindent By training a model ($\boldsymbol{\hat{\alpha}}_i,c_i$ pair) for each time series we can stack our fitted models to produce a functional representation analogous to Equation \ref{eq:var_lin}:
\begin{equation}
    \hat{Q}_{\boldsymbol{y}_t|\boldsymbol{y}_{t-1}}(q) =A_1\boldsymbol{y}_{t-1}+\boldsymbol{c}. \label{qvar}
\end{equation}

%\noindent The validation of individual links (\ie, the statistical significance of $A_i$) is comparatively more difficult in the quantile regression framework, compared to the traditional asymptotic results for the least squares estimator used in standard VAR. 
%nd here we follow the approach outlined in \cite{greene2012econometric}, which synthesizes findings from several studies \cite{tvals0, tvals1, rogers1993calculation} and proposes the subsequent solution:

\noindent Validating individual links in our estimated recurrence matrices $\hat{A}_1$ is relatively challenging for quantile regression, especially when compared to the conventional asymptotic results available for least squares estimators, such as those used in standard VAR. Our approach to this problem is to once against construct $t$ values associated with a Wald test, $t_{i,j}=\hat{A}_{i,j}/\hat{s}_{i,j}$. However, this necessitates an alternative estimate for the standard error $\hat{s}_{i,j}$. The distributional properties of quantile regression coefficients has been explored in numerous papers, and here we follow the approach outlined in \cite{greene2012econometric}, which summarises the works from several studies \cite{tvals0, tvals1, rogers1993calculation} and puts forward the following solution:
\begin{align}
    \text{Var}({\boldsymbol{\alpha}}_q)=(X'X)^{-1}X'DX(X'X)^{-1}, \label{eq:var}
\end{align}
where $D_i$ is a diagonal matrix containing elements:
\begin{align}
d_i = 
\begin{cases} 
\frac{q}{{f(0)}^2} & \text{if } y_t - \boldsymbol{\hat{\alpha}}^T X_t > 0 \\
\frac{1-q}{{f(0)}^2} & \text{otherwise},
\end{cases}
\end{align}
and $f(0)$ is the true probability density function of the disturbances %, $\epsilon_{t,q}$%=y_t-\hat{y}_{t,q}$
 evaluated at $0$. %This value is typically estimated using a single-parameter family of kernel density estimator 
Typically, this value is estimated using a single-parameter kernel density estimator \cite{greene2012econometric}:
\begin{align}
    \hat{f}(0)= \frac{1}{nh_{q}}\sum^T_{t=1} K\left[ \frac{\epsilon_{t,q}}{h_q} \right],
\end{align}

%\noindent For our study we utilise a Gaussian Kernel:
\noindent where $\epsilon_{t,q}=y_t-\hat{y}_{t,q}$. In our study we employ a Gaussian kernel for density estimation, which is defined as:
\begin{align}
K(x) = \frac{1}{\sqrt{2\pi}} e^{-\frac{x^2}{2}},
\end{align}

%\noindent Where we select a gaussian kernel, K, 

\noindent and select the bandwidth function, $h_q$, specified in \cite{hseather}. For each of the $i$ series, we then calculate $\hat{s}_{i,j}$ as the root of the $j$'th diagonal element of the Covariance matrix found in \autoref{eq:var}.

The kernel and bandwidth we select is available in numerous standard software packages. For our analysis, we make use of the Python package Statsmodels  \cite{seabold2010statsmodels}. This package is particularly efficient for our needs, as it solves %equation 
(\ref{eq:armin}) using recursive least squares --- an approach that scales effectively with the high number of covariates in our study. In addition to the median target $q=0.5$, we select a lower, $q=0.1$, and an upper quantile, $q=0.9$, as targets for this study. 

\subsection{Piecewise Quantile Regression}

To investigate whether there are unique causal effects in the tail regions of our input variables we extend our analysis beyond basic linear functions. A straightforward method to introduce nonlinear effects into %our current
a linear setup involves transforming the covariates, a process commonly known as basis expansion. One popular variant is the polynomial basis expansion, where both the raw covariate values and their higher-order transformations, such as squared returns, are taken into account. In our context, these transformations can be instrumental in capturing autoregression in the higher moments of the return series. For example, the effect of a covariate on the 90th percentile might not be linear with respect to $y_t$, but could be linear in $y^2_t$, suggesting a volatility shift akin to GARCH dynamics.

Yet, applying these methods to our data presents %certain
challenges: 
\begin{enumerate}
    \item The raw data %inherently
displays nontrivial kurtosis, and the introduction of squared and cubed values can exacerbate the associated estimation.
    \item Such embedding may yield results that aren't readily interpretable, e.g.  where the cumulative dependency of the outcome on the covariate is a complex%convoluted
     blend of the polynomial coefficients. For example, if the median of $y_t$ is positively influenced by $y_{t-1}$, negatively by $y^2_{t-1}$, and positively again by $y^3_{t-1}$, discerning an intuitive net effect becomes challenging.
\end{enumerate}

%Yet, applying these methods to our data presents %certain
%challenges: (1) The raw data %inherently
%displays nontrivial kurtosis, and the introduction of squared and %cubed values can exacerbates the associated estimation% errors %linked to this abnormality.
%(2) Such embedding may yield results that aren't readily %interpretable, e.g.  where the cumulative dependency of the %outcome on the covariate is a complex%convoluted
%blend of the polynomial coefficients. For example, if the median %of $y_t$ is positively influenced by $y_{t-1}$, negatively by $y^2_{t-1}$, and positively again by $y^3_{t-1}$, discerning an intuitive net effect becomes challenging.

%for instance, if median of $y_t$ positively depended on $y_{t-1}$, negatively on $y^2_{t-1}$ and positively on $y^3_{t-1}$ it would be difficult to provide an intuition around the net effect.

%Lastly, there is a relatively less straightforward relationship between the "tails" of the input and the target. We could consider the  Moreover, such a format could limit the representation of asymmetric or tail-specific causal effects, which are often explored in studies employing some form of asymmetric regression.

To circumvent these challenges, our solution utilises piecewise linear embeddings. This approach enables us to asymmetrically capture higher-order causal effects without compounding outlier estimation issues. The inherent linearity of this method, combined with the pinball loss function facilitates a relatively robust estimation of causal nonlinearities. The specific embedding we consider is a replacement of all instances of $X$ in the covariate equation with a 3-knot piecewise linear embedding:
%\begin{align}
%            X_i\rightarrow \{X_{i,L},X_{i,M},X_{i, U}\}= \%{X_{i}\times\mathcal{I}_{X_i<Q_{10}},X_{i},X_{i, U}\times\mathcal{I}_{X_i>Q_{90}}\}   
%\end{align}
%Where the middle embedding is unchanged, $X_i=X_{i,M}$, and the left and right embeddings are the multiplied by indicator functions for whether they are above or below the $10$ and $90$ quantile $Q_{90}$ for the covariate $X_i$. 
% \begin{align}
%             X_i\rightarrow \{X_{i,L},X_{i,M},X_{i, U}\}= \{\text{Min}(0, X_{i}-Q_{10}(X_i)),X_{i},\text{Max}(0, X_{i}-Q_{90}(X_i)\}   
% \end{align}
\begin{align}
            X_j\rightarrow \{X^-_{j},X_j,X^+_{j}\},
\end{align}
where: 
% \begin{align}
%     &X_{i,L} =  \text{Min}(0, X_{i}-Q_{10}(X_i)) \\
%     %&X_{i} = X_{i} \\
%     &X_{i, U} = \text{Max}(0, X_{i}-Q_{90}(X_i)).
% \end{align}
\begin{align}
    &X^-_{j} =  \text{min}(0, X_{j}-Q_{X_j}(0.1)) ,\\
    %&X_{i} = X_{i} \\
    &X^+_{j} = \text{max}(0, X_{j}-Q_{X_j}(0.9)).
\end{align}
The breakpoints for our piecewise embeddings ($X^-_{j},X_j,X^+_{j}$)  are selected as the unconditional quantiles of the input variables, $Q_{X_j}(0.1)$ \& $Q_{X_j}(0.9)$ to ensure that we consistently target tail effects across all coins. From this basis expansion, alongside the representation shown in \autoref{qvar} we can present our final model in the following format:
\begin{align}
    \hat{Q}_{\boldsymbol{y}_t|\boldsymbol{y}_{t-1}}(0.9) =A^-_{0.9}\boldsymbol{y}_{t-1}^-+A_{0.9}\boldsymbol{y}_{t-1}+A^+_{0.9}\boldsymbol{y}_{t-1}^+ +\boldsymbol{c}_{0.9} \label{eq:Q90},\\
    \hat{Q}_{\boldsymbol{y}_t|\boldsymbol{y}_{t-1}}(0.5) =A^-_{0.5}\boldsymbol{y}_{t-1}^-+A_{0.5}\boldsymbol{y}_{t-1}+A^+_{0.5}\boldsymbol{y}_{t-1}^+ +\boldsymbol{c}_{0.5}\label{eq:Q50},\\
    \hat{Q}_{\boldsymbol{y}_t|\boldsymbol{y}_{t-1}}(0.1) =A^-_{0.1}\boldsymbol{y}_{t-1}^-+A_{0.1}\boldsymbol{y}_{t-1}+A^+_{0.1}\boldsymbol{y}_{t-1}^+ +\boldsymbol{c}_{0.1}.\label{eq:Q10}
\end{align}

%^\noindent In conclusion, t
\noindent This technique allows us to produce 9 separate networks, each associated with one of the $A^*_q$ and designed to locally capture the causal structure in some sub-region of the input and output densities. The interpretation of our sub-networks are relatively intuitive, for instance, a link $i\rightarrow j$ in $A^+_{0.9}$ denotes the effect that large observations of asset $i$ have on the upper quantiles of $j$. Similarly, a link in $A^-_{0.1}$ indicates the low observations of asset $i$  effects the lower quantiles of asset $j$.

To obtain coefficients that match our intended interpretation of links we make several minor structural modifications, such that the model space and parameter interpretation are changed without effecting the actual fitted curves. %(1)
First, we shift the model space for the $q=0.1$ and $q=0.9$ regression from modelling the raw target quantiles, to fitting the quantiles of the residual from the median regression: $\epsilon_t=y_t-\hat{y}_{t,0.5}$. This changes  \autoref{eq:Q90} to the following: 
\begin{align}
    \hat{Q}_{\boldsymbol{y}_t|\boldsymbol{y}_{t-1}}(0.9)&=\hat{Q}_{\boldsymbol{y}_t|\boldsymbol{y}_{t-1}}(0.5) + \hat{Q}_{\boldsymbol{\epsilon}_t|\boldsymbol{y}_{t-1}}(0.9),\\
    &= \hat{y}_{t,0.5}+ A^-_{0.9}\boldsymbol{y}_{t-1}^-+A_{0.9}\boldsymbol{y}_{t-1}+A^+_{0.9}\boldsymbol{y}_{t-1}^+ +\boldsymbol{c}_{0.9}
\end{align}
This allows us to separate the scale and location effects of our target variables, changing the focus of the $Q(0.1)$ and $Q(0.9)$ targets towards describing the risk and volatility effects in the influence structure. For instance, a link in $A^+_{0.9}$ now denotes that the distance of the right tail from the median is being effected by $\boldsymbol{y}_{t-1}$. In the prior specification, we would have a link in $A^+_{0.9}$ if the entire density shifted, even if there was no tail-specific changes. 
%(2) 

Second, when evaluating our final networks we change the tail-input subnetworks to a format that represents the net effect, rather than the partial effects originally present in Equations \ref{eq:Q90}, \ref{eq:Q50} and \ref{eq:Q10}. %To do so we analyse the `net slope' coefficient matrices, $\mathbb{A}^\pm_{q}\boldsymbol{y}=(A^\pm_{q}\boldsymbol{y}+A_{q}\boldsymbol{y})$,
To do so we analyse the `net slope' coefficient matrices, $\mathbb{A}^\pm_{q}=(A^\pm_{q}+A_{q})$, with the covariances adjusted to reflect this sum:
\begin{align}
    \text{Var}(\mathbb{A} ^\pm_{q})=\text{Var}(A^\pm_{q})+\text{Var}(A_{q})-2\text{Cov}(A^\pm_{q},A_{q}).
\end{align}
%where the covariance term can be obtained from the off diagonals of \ref{eq:var}

%e $A^*_q$. It is in this way that we break apart the dependencies into their sections, with for instance, a link $i\rightarrow j$ in $A^+_{90}$ denoting the effect that large observations of $X_i$ has on the upper quantiles of $X_j$.

%\noindent In this study we are particularly concerned with 3 target quantiles q, namely, the median (q=50), and the 10th and 90th quantiles. This allows us to capture location and scale information from the conditional density $F_{\boldsymbol{y}_t|\boldsymbol{y}_{t-1}}$, while minimising the number of parameters. A requirement necessitated by our piecewise linear basis expansion, leading us to have 783 covariance for our 8000 observations (our ratio n/p is around the lower bounds of the colloquially accepted requirement). 

\begin{figure}[ht!]
    \centering
    \begin{subfigure}[t]{0.32\textwidth}
        \centering
        \includegraphics[width=\textwidth]{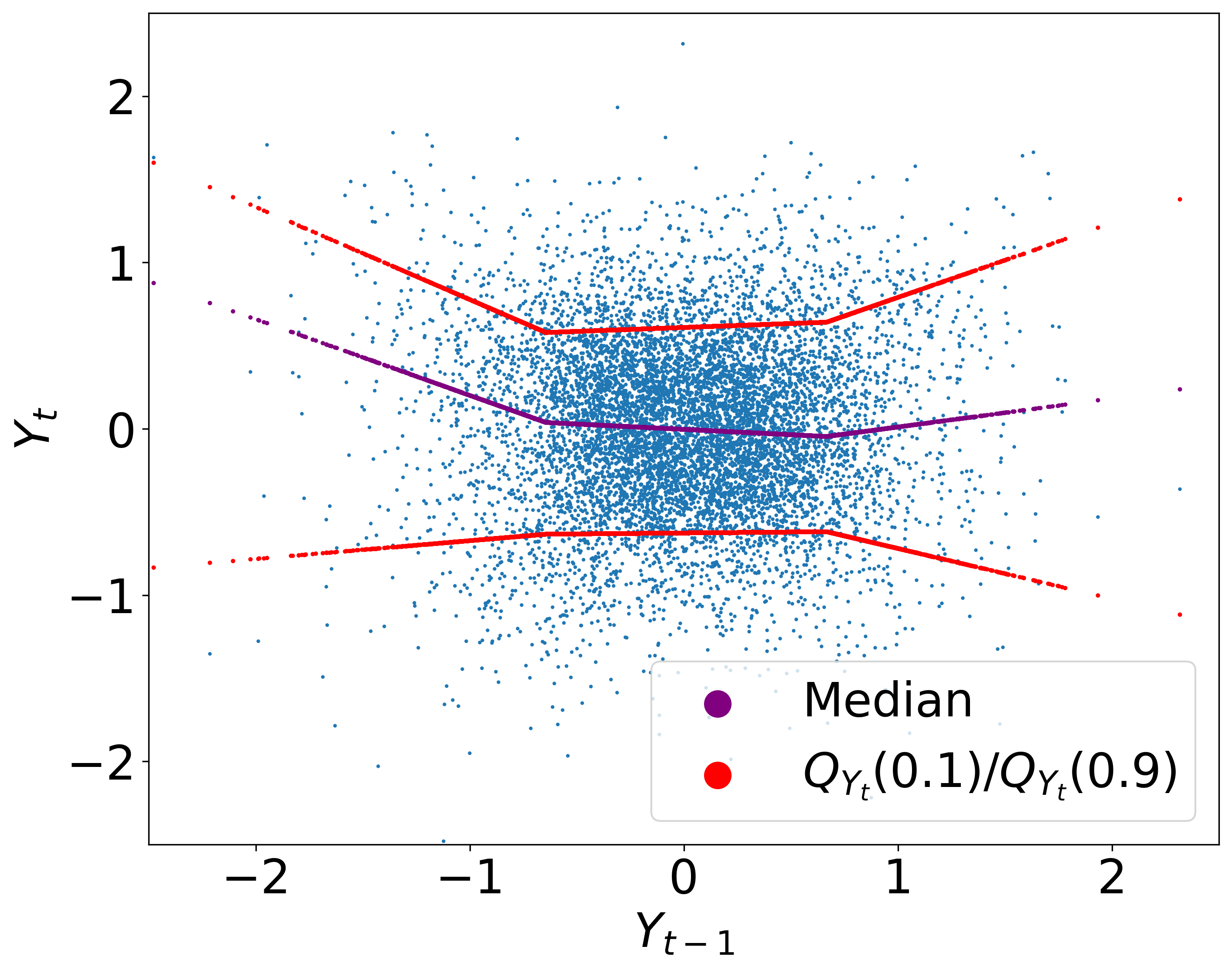} % Replace with your image filename
        \caption{Bitcoin self-causal effects.}
        \label{fig:btc_self}
    \end{subfigure}
    \hfill
    \begin{subfigure}[t]{0.32\textwidth}
        \centering
        \includegraphics[width=\textwidth]{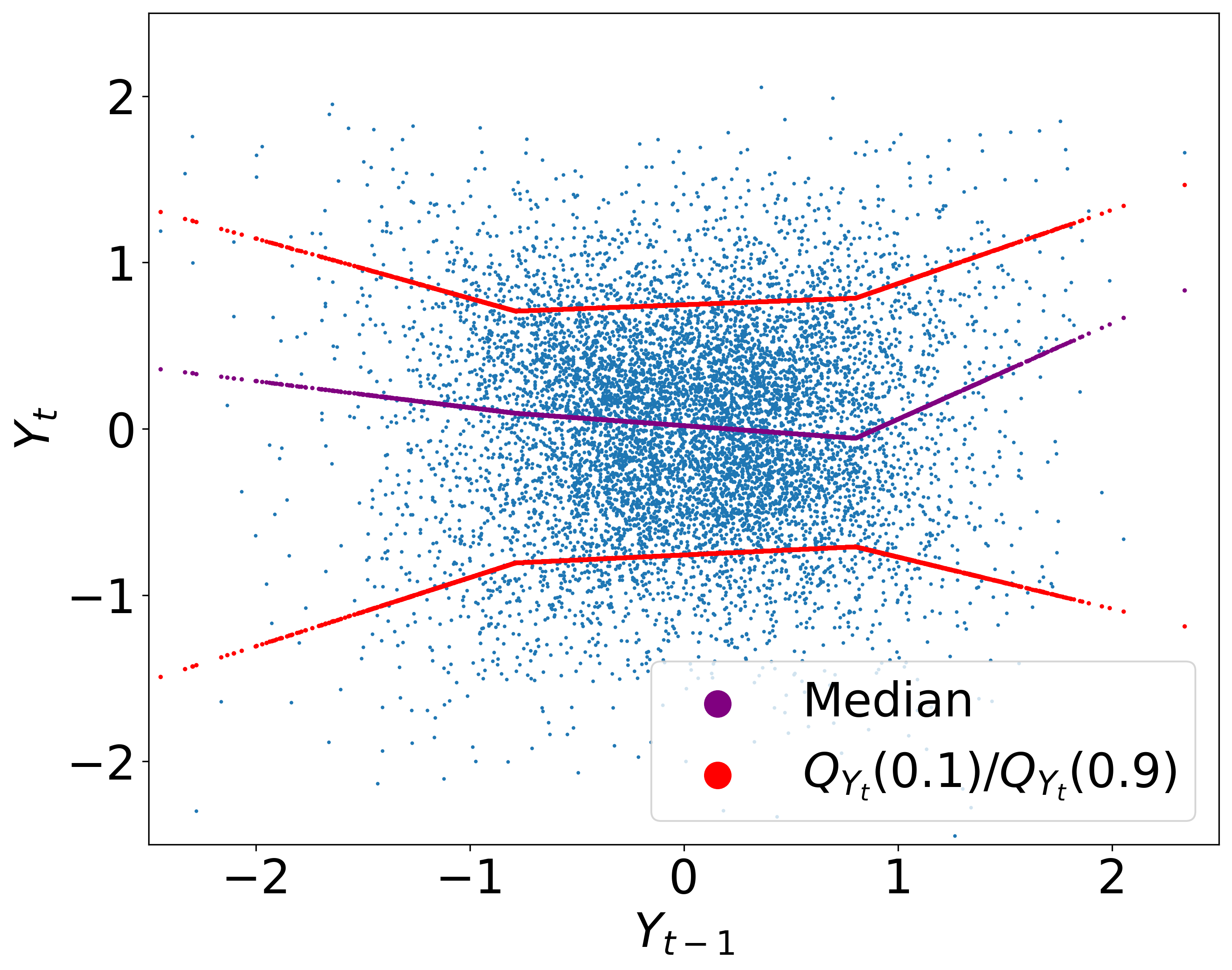} % Replace with your image filename
        \caption{Ethereum self-causal effects.}
        \label{fig:eth_self}
    \end{subfigure}
    \hfill
    \begin{subfigure}[t]{0.32\textwidth}
        \centering
        \includegraphics[width=\textwidth]{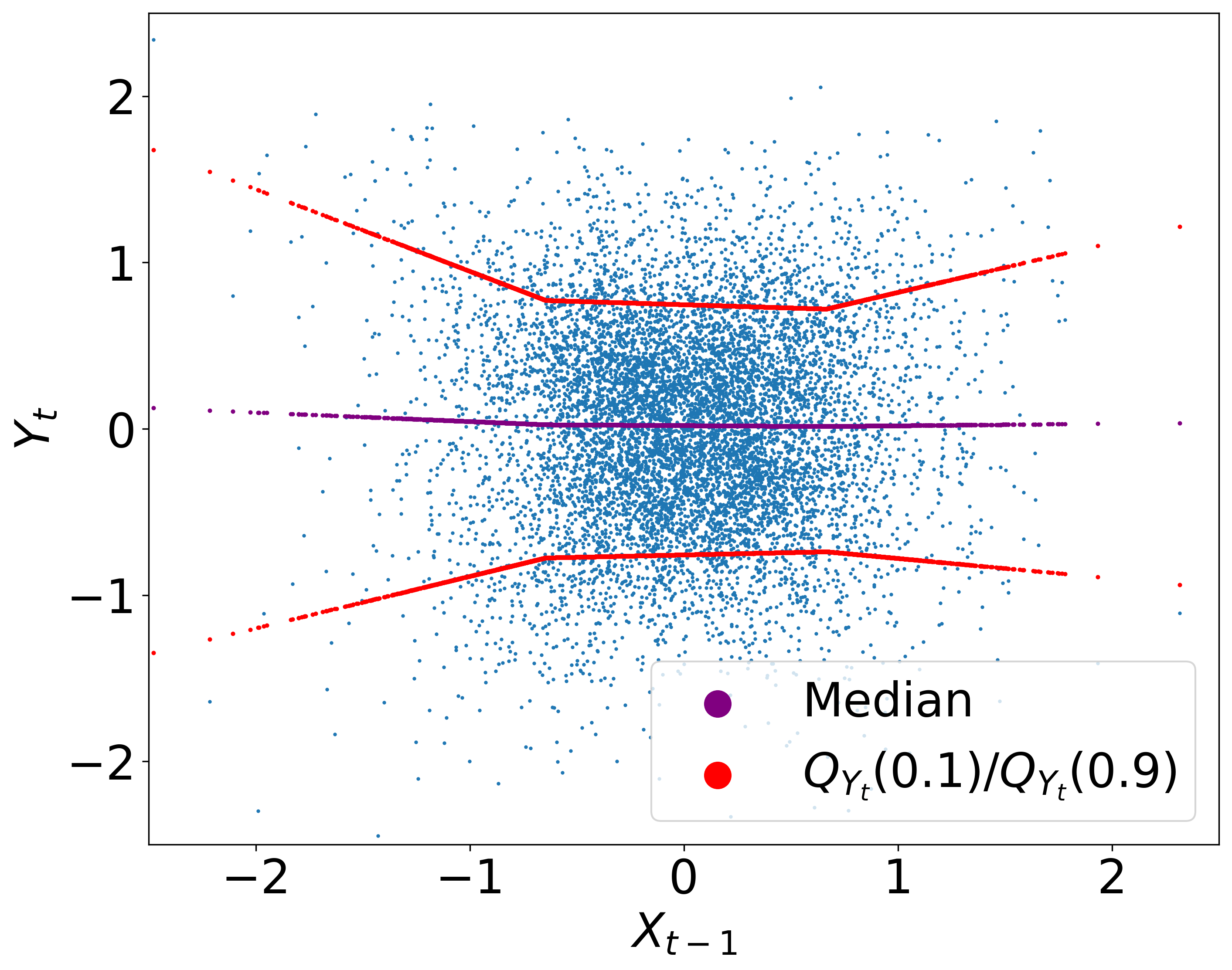} % Replace with your image filename
        \caption{Bitcoin $\Rightarrow$ Ethereum cross-causal effects.}
        \label{fig:btc_eth_cross}
    \end{subfigure}
    \caption{Fitted causal effects within a 2 node Bitcoin $\Leftrightarrow$ Ethereum example network. For each data point along the horizontal axis we display the predicted target quantiles, allowing us to visualise the type of relationships P-QVAR identifies.}
    \label{fig:fitted_effects}
\end{figure}
\vspace{-10mm}

\newpage
\subsection{Graphical Representations:}
While \autoref{eq:Q10}, \autoref{eq:Q50} and \autoref{eq:Q90} may appear mathematically esoteric, the fitted relationships are quite visually intuitive. \autoref{fig:fitted_effects} provides a visual indication of the relationships that P-QVAR can fit, showing variable effects estimated from an example network containing only Bitcoin (BTC) and Ethereum (ETH). This simple example demonstrates several archetypal relationships that motivate our design: 
\begin{itemize}
    \item We observe in \autoref{fig:btc_self} and \autoref{fig:eth_self} that that the median response often has a differential, larger impact for tail events.
    \item The upper and lower quantiles tend to systematically display a double-cone, or hourglass shape, after an intermediate period of low dependence. This pattern is consistent with autoregression in the volatility of returns (consistent with GARCH models). 
    \item \autoref{fig:btc_eth_cross} demonstrates that even when the variables appear independent in terms of central tendency (median), they may have relationships occurring for the outer tails, indicating risk-spillover. 
\end{itemize}
The nonlinear profile for all targets motivates our piece-wise embeddings, ${X^+_i, X^-_i}$, while the differing dependence across our target regions drives us to asses multiple quantiles.

%We see the archetypal mutual excitation patterns across the quantiles, alongside the generally idiosyncratic median response effects. 

%We also strive to visualise the overall effects observed from an entire set of sub-networks, $$.

\begin{figure}[htp!]

\centering

% First Row
\begin{subfigure}[t]{.475\textwidth}
  \centering
  % include first image
  \includegraphics[width=\linewidth]{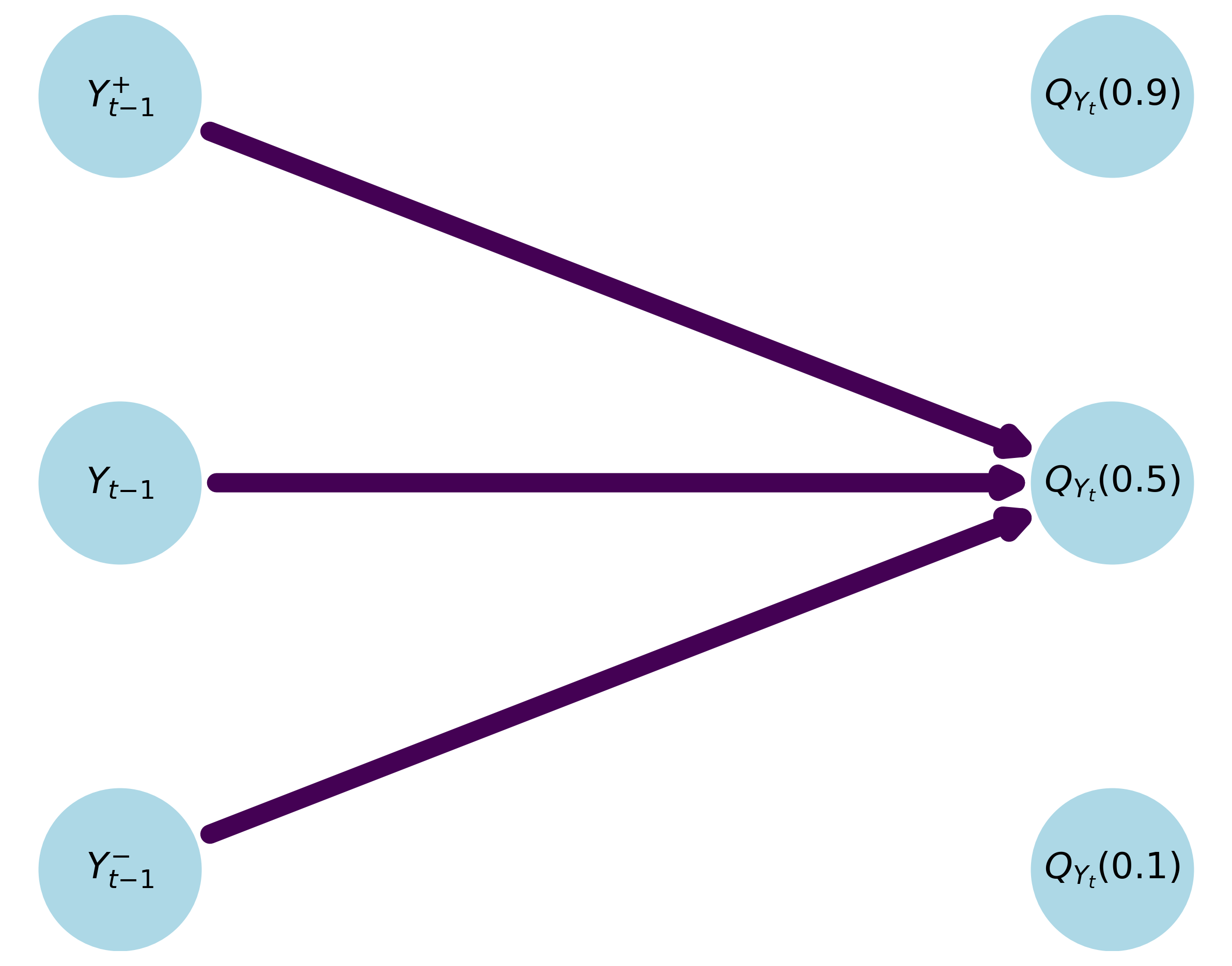}  
  %\caption{VAR like behaviour is shown with consistent median-response effects across all input regions. The negative effect visible from the link coloration indicates  }
   \caption{VAR behaviour is visible as consistent median-response links across all input regions. }
  \label{fig:var_arch}
\end{subfigure}%
\hfill
\begin{subfigure}[t]{.475\textwidth}
  \centering
  % include second image
  \includegraphics[width=\linewidth]{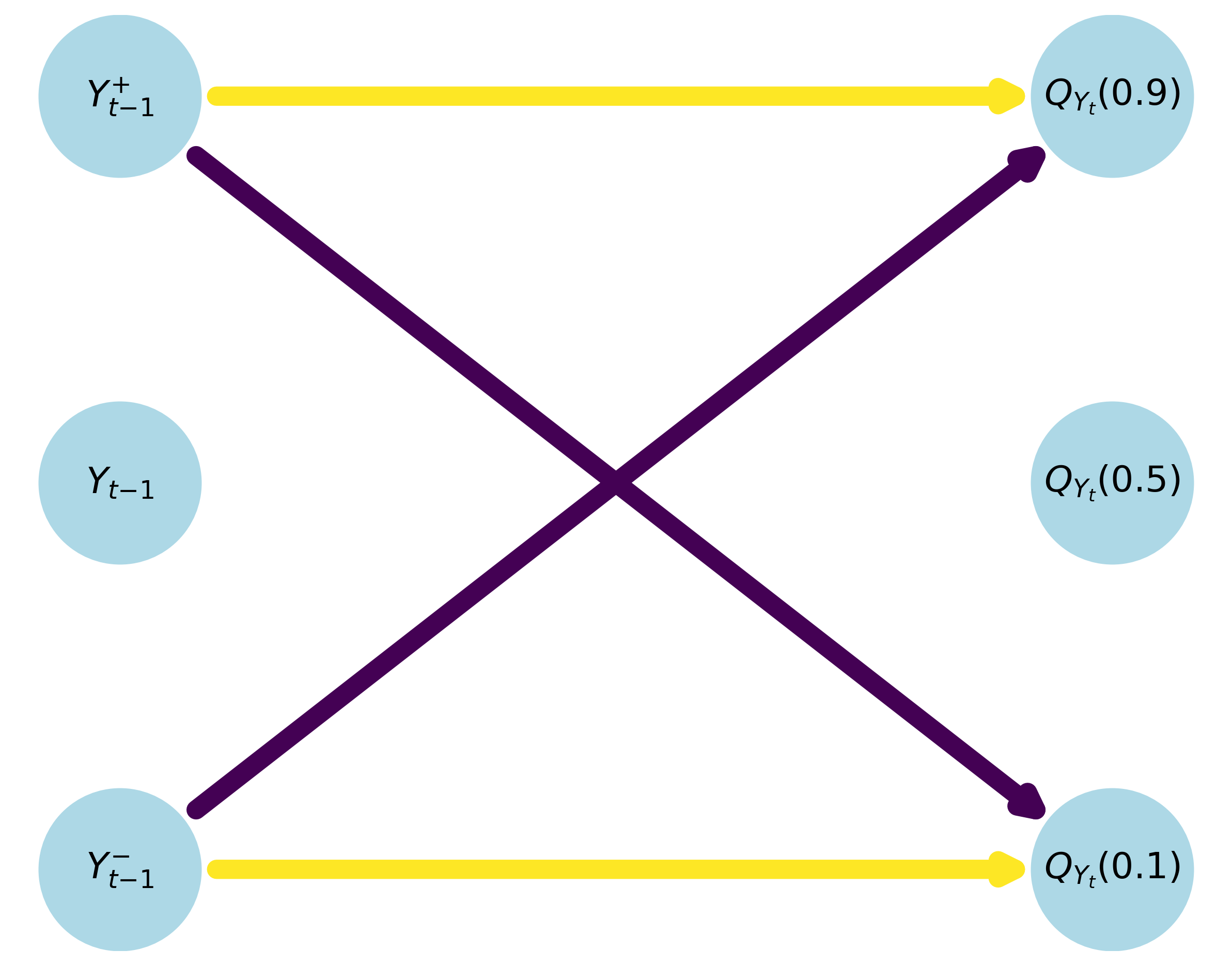}  
  \caption{GARCH type effects generate positive links between matching tails, with negative effects for opposing pairs and an absence of median input/target link.}
  \label{fig:garch_arch}
\end{subfigure}

% Second Row
\begin{subfigure}[t]{.475\textwidth}
  \centering
  % include third image
  \includegraphics[width=\linewidth]{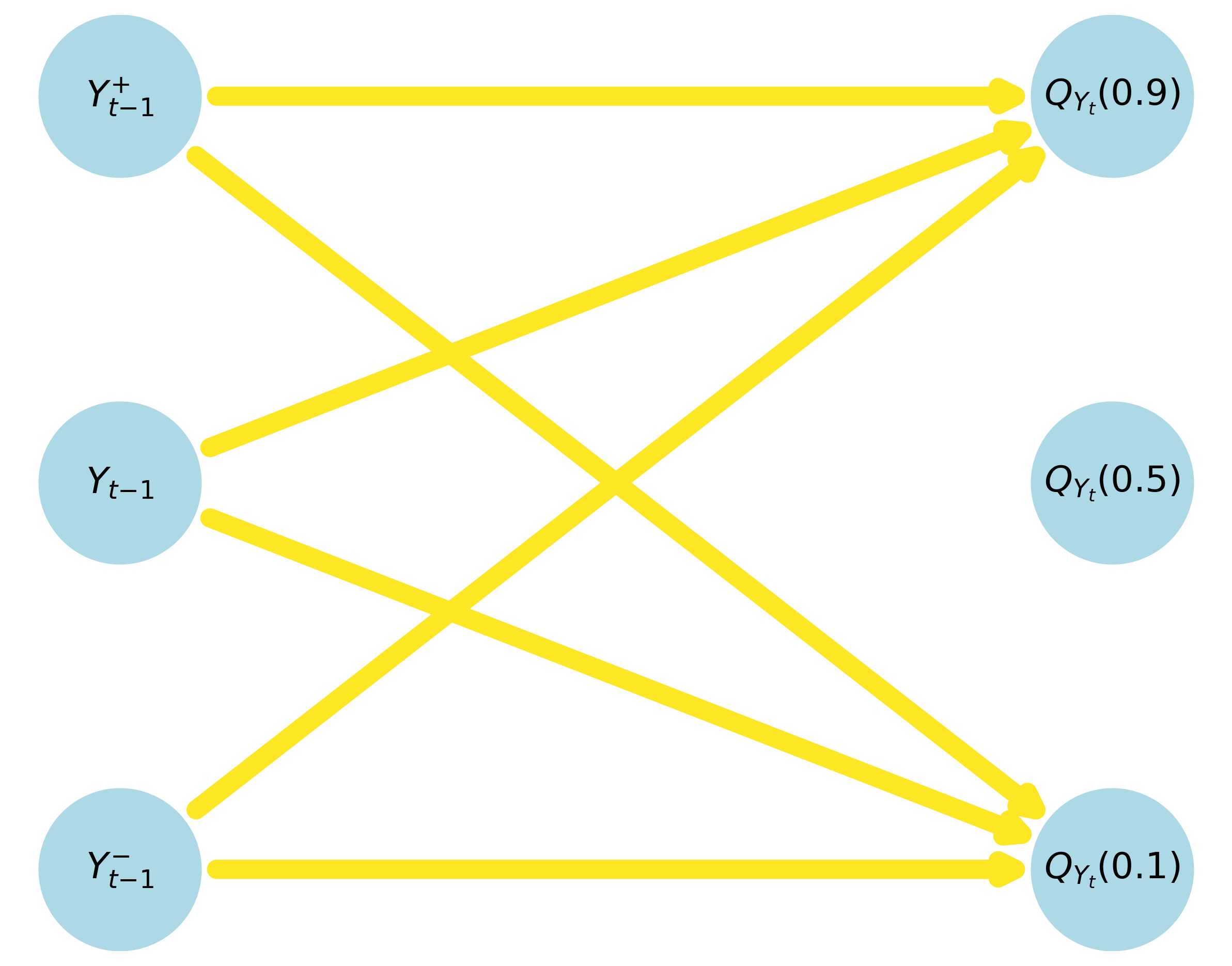}  
  \caption{Skew autoregression effects would show positive links between all input regions towards the tails.}
  \label{fig:skew_arch}
\end{subfigure}%
\hfill
\begin{subfigure}[t]{.475\textwidth}
  \centering
  % include fourth image
  \includegraphics[width=\linewidth]{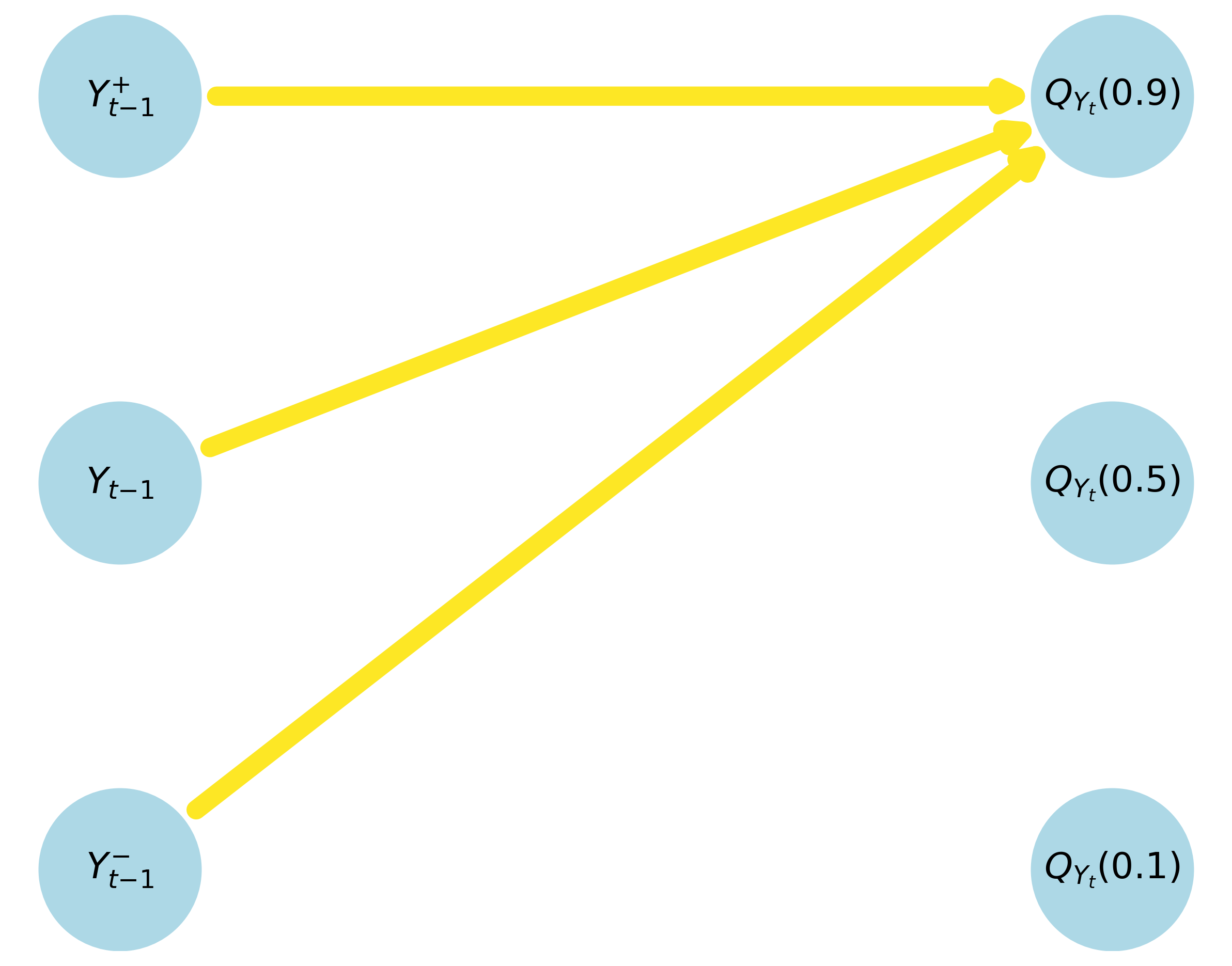}  
  \caption{Asymmetric effects such as consistent positive response towards one tail can be represented with P-QVAR, despite difficulties identifying these effects in standard moment based approaches.}
  \label{fig:mixed_arch}
\end{subfigure}
\vspace{5mm}
\caption{Quantile Influence Graphs for several %types
examples of standard distributional relationships. Edge coloration indicates the direction of influence, with yellow lines denoting a positive effect, and purple lines indicating a negative effect.}
\label{fig:example_QIG}

\end{figure}

In order to visually asses the types of causal effects identified across the entire set of sub-networks, $A^\pm_i$, we create a Quantile Influence Graph (QIG). This visualisation contains nodes for all input and target regions that we consider, with the edges corresponding to the different sub networks, $A^\pm_i$. To demonstrate our QIG, \autoref{fig:example_QIG} displays the type of structures that would be identified by running our method on several archetypal models. Specifically, we visualise autoregression in increasing moments, starting with a negative standard VAR, then GARCH effects ($\sim$ARMA of volatility) and lastly show autoregression of the skew, followed by an assymetric effect.

%\section{Results}
\section{Empirical Study}
\label{sec:results}
\subsection{Data}
\label{sec:Data}

% {\bf MR: Throw in a citation to your preprint for the other paper, saying something like "for more details see \cite{}}
\vspace{-1mm}
% Our dataset contains hourly prices from 260 cryptocurrencies from 1/1/2021 to 1/1/2022, along with their market capitalisation (June 2022 figures). The selected currencies were derived from the 750 coins with highest capitalization at time of collection (June 2022). However, a significant number of these, mainly those with lower capitalization, had incomplete or missing price histories and had to be excluded. The resulting data contains $79.6\%$ (836B of 1.05T) of the total capitalisation of the market. The returns $y_t$ are generated by taking the logged ratio of subsequent observations in the original price series $p_t$, quoted in terms of the Coin/USD relation: 
Our dataset comprises hourly prices for 260 cryptocurrencies spanning from January 1, 2021, to January 1, 2022, along with their market capitalizations as of June 2022. The selected currencies were drawn from the top 750 cryptocurrencies by market capitalization at the time of data collection (June 2022). However, a substantial portion of these, primarily those with lower capitalizations, were excluded due to incomplete or missing price histories. The final dataset represents 79.6\% of the total market capitalization (836B out of 1.05T). The returns, $y_t$, are computed as the logged ratio of consecutive observations in the original price series, 
$p_t$, expressed in terms of the Coin/USD pair:
%. Specifically, the hourly returns are calculated as 
 \ie $y_t = 100\log\left(p_t/p_{t-1}\right)$. 
For comprehensive details on this dataset see \cite{cornell2023vector}. To minimise the effects of data artifacts, such as coordinated pump and dump schemes, we take a stabilising transformation $\text{sign}(y_t)\times \log(|y_t|+1)$. Almost all hourly returns are in the low single digit $\%$ region, where this transform is near-linear, however it flattens the infrequent events where we may see some of the low valuation coins have high double, or even triple digit returns before reverting in the next period. 

%which occur infrequently and os

%For our purposes, the key takeaway is that our dataset shows high levels of capitalization dependent non-gaussianity.

\vspace{-3mm}

%\newpage
%\subsection{Self-effect analysis}
%\subsection{Analysis of Self-Causality}
\newpage
\subsection{Self-Causality}

%Given the particularly strong presencer of self-effects in causal networks for both mean-response and volatility effects we wish to pay particular attention to these effects. 

\begin{figure*}[h!]
     \centering
         \includegraphics[width=\textwidth]{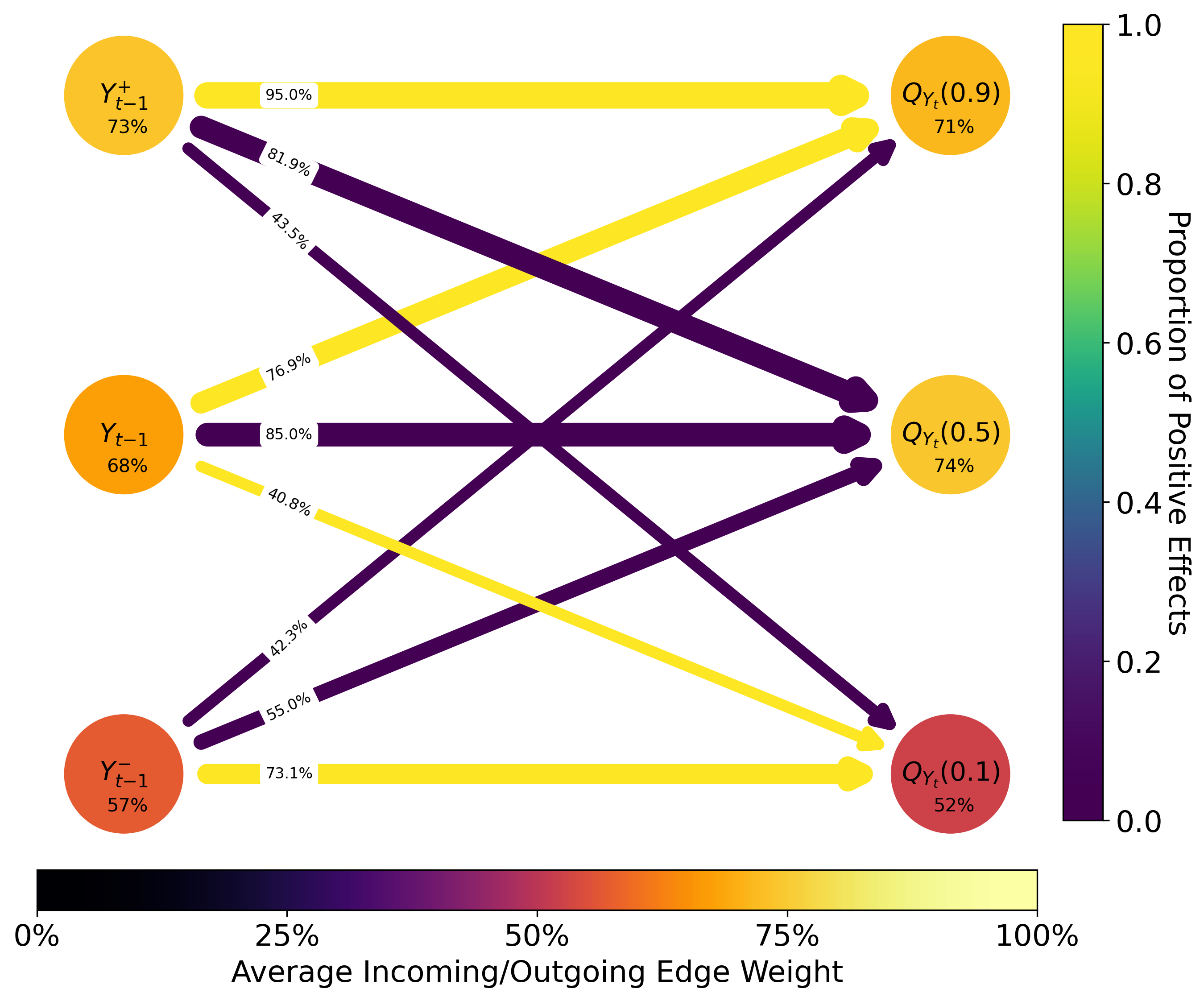}
         %\vspace{-1cm}
         %\caption{Node-based representation of self-causal effects. Edge weights indicate \% of possible links present, with coloration indicating effect sign propensity. }
         \caption{%Node-based representation 
         Quantile Influence Graph of self-causal effects. Edge weights indicate the proportion of possible links present, with coloration indicating effect sign propensity. Node coloration (and weights) indicates the average outgoing/incoming edge weight. The self-causal effects in our data display high directional regularity, with mean reversion, mutual excitation between the tails, and an asymmetric transfer of skew.}
         \label{fig:self_edges}
\end{figure*}

\autoref{fig:self_edges} presents a QIG of the self-edge effects (\ie the diagonal elements of $A^*_i$) in our overall causal network. In general, there is a high proportion of potential self-links that are statistically significant (65.8\% of possible links), exhibiting pronounced directional trends. More specifically:

\begin{itemize}
    \item[\textbullet] We observe pronounced mean-reversion effects for the median return target, similar to the negative VAR example shown in \autoref{fig:var_arch}:
    \begin{itemize}
        \item[-] This is particularly evident for the Linear-Median link, with 85\% of coins exhibiting this effect, and 99.1\% of these links being negatively signed. The right-tail to median link has equivalent behaviour, with almost all (99.5\%) of the 81.9\% of identified links being negatively signed. 
        \item[-] This effect is diminished in the negative-tail input region, with only 55\% of coins showing significant negative-tail price reversion (97.9\% negative).
        
    \end{itemize}
    \item[\textbullet] The tail-tail edge effects align with excitatory or GARCH-like behaviour (such as \autoref{fig:garch_arch}), further showing asymmetries that suggest an imbalanced transfer of Skew.
    \begin{itemize}
        \item[-] The Upper-Upper link is the strongest of all effect, with 95\% of possible links being present --- all of which are positively signed. The Lower-Lower link is also prominent but slightly reduced, with 73.1\% of links being significant, and 99.4\% positive effects.
        \item[-] Both the Lower-Upper and Upper-Lower links are relatively weaker, with symmetric effects (43.5\% and 42.3\%), and primarily negative signs (96.3\% and 96.4\%).
    \end{itemize} 
    
    \item[\textbullet] The overall node weight asymmetries suggest that negative events have a weaker effect in terms of both incoming (73\% vs 57\%), and outgoing influence (71\% vs 52\%). The strongest outgoing effects stem from large positive events, with the median target having the highest count of incoming influences. 
    \vspace{0.5cm}
    
    %Overall, we conclude that there are high levels of self causality (xyz\% of possible links) in cryptocurrency data, and that these relationships tend to be complex, asymmetric and tail focused. 

    %Overall, we conclude that there are high levels of complex, asymmetric and tail focused self causality in cryptocurrency markets. 

    %Overall, we conclude that there are high levels of self-causality in our cryptocurrency market data. The observed causal structures tend to be very prototypical, with almost all links having the same directional trends. This relationship is characterised by mean reversion in the density centre, and mutual excitation between the tails. We also observe notably weaker dependence for negative tail events. This combination of effects indicates quite erratic price behaviour, implying that we see generally reverting, self-exciting swings that have a propensity to enter short lived bull-runs. 
\end{itemize}

\noindent Overall, we conclude that our cryptocurrency market data exhibits high levels of self-causality. The observed causal structures tend to be highly prototypical, with nearly all links displaying the same directional trends. This relationship is characterized by mean reversion at the density center and mutual excitation between the tails. Additionally, we observe notably weaker dependence for negative tail events. This combination of effects suggests erratic price behavior, implying generally reverting, self-exciting swings with a tendency to enter short-lived bull runs.

    %financial context? applications? comment on market efficiency perhaps.

%\item[-] The Linear-Upper and Linear-Lower effects are consistent with skew transfernce (are they though????. 

%\newpage
%\subsection{Cross effect analysis}
%\subsection{Analysis of Cross-Causality}
\subsection{Cross-Causality}

\begin{figure*}[h]
     \centering
         \includegraphics[width=\textwidth]{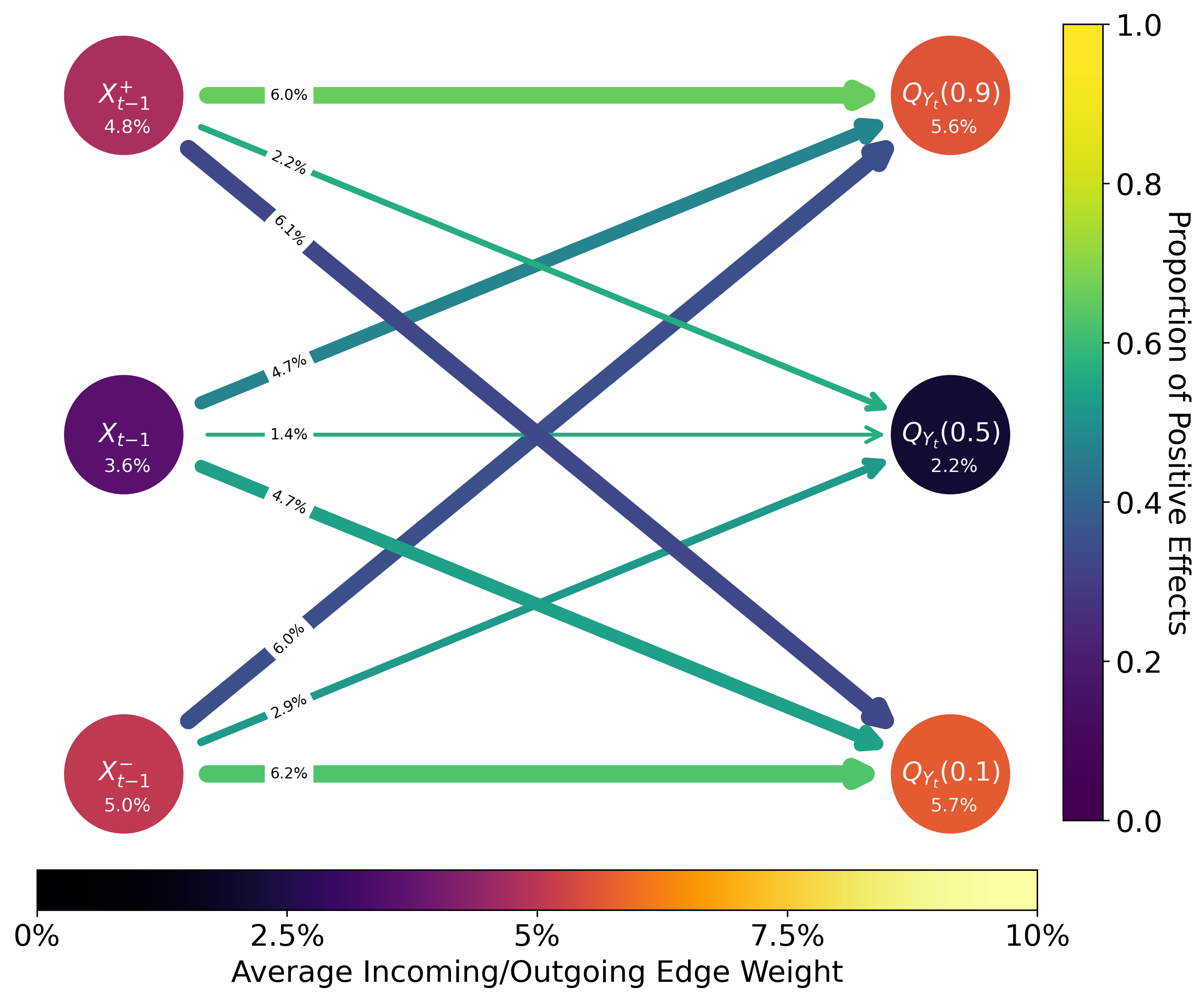}
         %\vspace{-1cm}
         %\caption{Node-based representation of cross-causal effects. Edge weights indicate \% of possible links present, with coloration indicating effect sign propensity. }
         \caption{%Node-based representation
         Quantile Influence Graph of cross-causal effects. Edge weights indicate the proportion of possible links present, with coloration indicating effect sign propensity. Node coloration (and weights) indicates the average outgoing/incoming edge weight. The cross-causal effects display a highly symmetric profile, with a focus on tail-tail interactions and an absence of strong directional trends.}
         \label{fig:cross_edges}
\end{figure*}

Figure \ref{fig:cross_edges} presents the same graphical representation as in Figure \ref{fig:self_edges}, but now for cross-causal effects (\ie off-diagonal elements of $A^*_i$). Generally, there is a greatly reduced number of statistically significant effects (4.5\% of possible links), with a comparatively high degree of symmetry. Further: 

\begin{itemize}
    \item[\textbullet] The median target now displays the lowest average incoming effect (2.2\% of possible links), with the strongest influence coming from tail events, rather Linear-Median response.
    \begin{itemize}
        \item[-] The most pronounced asymmetry in cross-causal effects is the heightened presence of $Q(0.1)$-median links, standing at 2.9\%, compared to 2.2\% for the $Q(0.9)$-median effects. 
    \end{itemize}
    \item[\textbullet] While tail-tail effects have also diminished, their reduction is less pronounced, leaving them as the most prominent remaining effect.
    \begin{itemize}
        \item[-] We no longer observe such asymmetry in these effects, with the matching tails and opposite tails all having approximately similar frequency ($\approx6\%$), with signage patterns slightly favoring excitatory over anti-excitatory behavior.
    \end{itemize} 
    \item[\textbullet] The directionality of edges remains closer to 50\% across all edge types, suggesting that cross-causal effects are less proto-typical and show diverse influences.
    \vspace{0.5cm}
    %\item[\textbullet] The overall node weights are relatively symmetric, indicating that both large positive and negative events 
    %exert approximately equal causal impact within the network. generally, there is more pronounced cross-causality between the outer quantiles than between the linear input or the median response aspects of the causal network.

\end{itemize}

\noindent We conclude that there is comparatively limited, but still present causality between coins in the cryptocurrency market. These relationships generally occur between the tails and often reflect excitatory patterns. The effects are quite symmetric, indicating that for cross-causal effects, negative and positive tail events have similar levels of influence. In the financial context, this implies that while there may be limited return spillover (linear-median links), the volatility and risk of coins may rapidly propagate through complex causal chains. For market participants, this indicates that cross-causal effects are relevant for risk management, but might have less application to statistical arbitrage style trading.

%This suggests that cross-causal effects 

%In the financial context, these results are quite intuitive. The presence of arbitrage trading would be expected to minimise the presence of direct return spillover (linear-median links)

%\subsection{CCDFS}
%\subsection{Analysis of Network Structure}
\subsection{Network Structure}

\begin{table}
    \centering
    %\caption{Summary statistics for all causal sub-networks. Displaying mean and standard deviation, $\sigma$, of varying attributes. $\rho$ denotes the correlation of said attribute to market capitalisation, with stars denoting the statistical significance. P(Down) denotes the probability that the source of a link has higher capitalisation than the target.}

    \caption{Summary statistics for all causal sub-networks. Displaying mean and standard deviation, $\sigma$, of nodes degrees, along with the Spearman rank correlation ($\rho$) with market capitalisation and the statistical significance of independence from capitalization, with significance levels indicated by asterisks($*\rightarrow p\leq0.05, **\rightarrow p\leq0.01, ***\rightarrow p\leq0.001$). P(Down) denotes the probability that the source of a link has higher capitalisation than the target.}

    \label{tb:3x3}
    % Row 1
    \begin{subtable}{0.32\textwidth}
        \centering
        \input{UL.tex}
        \vspace{-4mm}
        \caption{Left Tail - $Q(0.9)$}
    \end{subtable}
    \hfill
    \begin{subtable}{0.32\textwidth}
        \centering
        \input{UM.tex}
        \vspace{-4mm}
        \caption{Linear-$Q(0.9)$}
    \end{subtable}
    \hfill
    \begin{subtable}{0.32\textwidth}
        \centering
        \input{UR.tex}
        \vspace{-4mm}
        \caption{Right Tail - $Q(0.9)$}
    \end{subtable}

    % Row 2
    \begin{subtable}{0.32\textwidth}
        \centering
        \input{ML.tex}
        \vspace{-4mm}
        \caption{Left Tail - Median}
    \end{subtable}
    \hfill
    \begin{subtable}{0.32\textwidth}
        \centering
        \input{MM.tex}
        \vspace{-4mm}
        \caption{Linear - Median}
    \end{subtable}
    \hfill
    \begin{subtable}{0.32\textwidth}
        \centering
        \input{MR.tex}
        \vspace{-4mm}
        \caption{Right Tail - Median}
    \end{subtable}

    % Row 3
    \begin{subtable}{0.32\textwidth}
        \centering
        \input{LL.tex}
        \vspace{-4mm}
        \caption{Left Tail - $Q(0.1)$}
    \end{subtable}
    \hfill
    \begin{subtable}{0.32\textwidth}
        \centering
        \input{LM.tex}
        \vspace{-4mm}
        \caption{Linear - $Q(0.1)$}
    \end{subtable}
    \hfill
    \begin{subtable}{0.32\textwidth}
        \centering
        \input{LR.tex}
        \vspace{-4mm}
        \caption{Right Tail - $Q(0.1)$}
    \end{subtable}
\end{table}

\begin{figure*}[ht!]
     \centering
         \includegraphics[width=\textwidth]{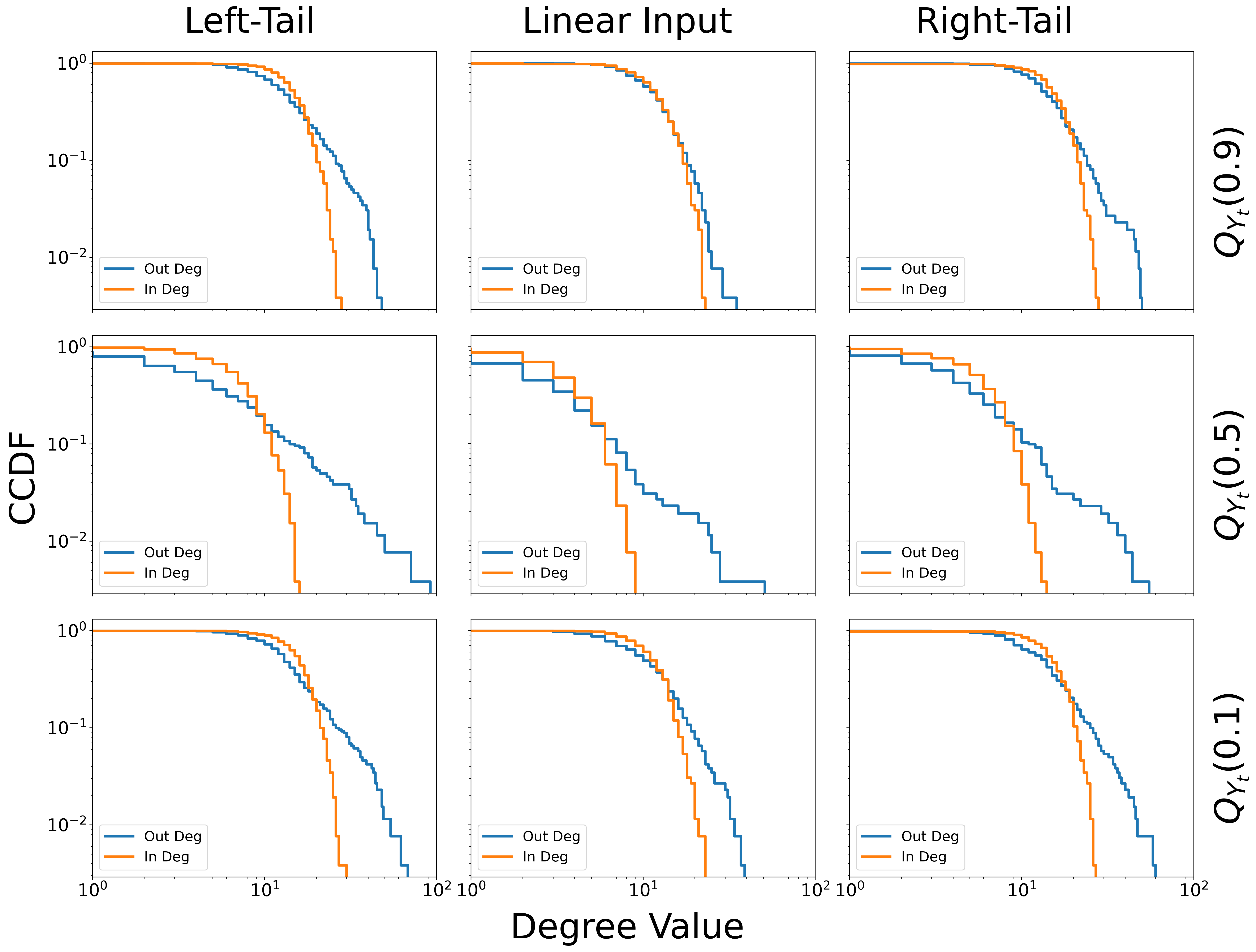}
         %\vspace{-1cm}
         \caption{Log-Log scale complementary cumulative distribution functions for all 9 sub graphs, showing both in and out degree forms. The columns separate source regions, while rows denote the targets. Linear decay in these survival functions is consistent with fat-tailed distributions of influence.}
         \label{fig:3x3CCDF}
\end{figure*}

Figure \ref{fig:3x3CCDF} displays the log-log Complementary Cumulative Distribution Functions (CCDF) for the degree distributions of our 9 sub-networks. We see that across all forms of sub network, the out-degree distributions display comparatively heavier tails, indicating that out-degree, the measure of external influence is more concentrated than incoming influence. This is particularly pronounced for the median response targets, where we see quite slow decay in the survival function. Across all targets, there appears to be higher scaling for the tail-effect distributions, with the largest amount of scaling coming from the left-tail to median response type link.

%For both the left and central (linear) covariate effects of the outer quantiles we see relatively lower scaling, with elevated outliers effect for both residual quantiles. 

%we observe that there is a significant overall propensity for the out-degrees to show a higher degree of scaling and distributional skew, with a larger outlier focus. This is particularly pronounced fro the median response targets, where we see quite slow decay in the survival function. 

%To further investigate the concentration of incoming and outgoing influence we can consider the lorenz curves and Gini co-efficient for each of these degree distributions. 

\vspace{3mm}
\noindent Table \ref{tb:3x3} presents the summary statistics for our individual causal subnetworks, displaying the mean, standard deviation, and correlation to market capitalization for several attributes.

\begin{itemize}
    \item[\textbullet] Statistically significant correlations are observed between out-degree and market capitalisation for both input tail-effects directed towards the median and $Q(0.1)$ targets. Notably, these correlations are positive for all tail effects except for the right-tail to $Q(0.9)$ link.

    %\item[\textbullet] We observe positively signed, statistically significant correlations between In-Degree and market capitalisation for both matching tail subnetworks. and a negatively signed signficant correlation for the linear-median response. 

    \item[\textbullet] For in-degree, there are positively signed, statistically significant correlations with market capitalisation for both matching tail subnetworks (left-tail to $Q(0.1)$, right-tail to $Q(0.9)$). Conversely, a statistically significant negative correlation is discerned for the linear-median response.

    %\item[\textbullet] We observe that links involved in both the upper and lower residual subnetworks have a slight proclivity towards pointing 'downstream' in terms of their market capitilisation. 

    %\item[\textbullet] Links within both the upper and lower residual subnetworks demonstrate a slight inclination to point 'downstream' in terms of their market capitalisation. This effect is exacerbated for the median response networks, where we see up to $\approx80\% $of links pointing downstream in the Linear-Median subnetwork. 

    \item[\textbullet] Median response links have a mild tendency to point to point downstream towards lower-valued coins (52-55\%). %in terms of their market capitalisation (52-55\%). 
    This effect is reversed for the quantile targets, with a slight bias to upstream links. 

    \item[\textbullet] The standard deviation for out-degree and in-degree corroborates the observations for the CCDFs. Outgoing influence is relatively disperse, with some coins being influence drivers, while others have very low impact. Comparatively, in-degree is quite regular, with most coins having similar amounts of in-coming influence.

    %This tendency is particularly pronounced within the median response networks. In the linear-median response subnetwork, an overwhelming majority, approximately 80%, of links are directed towards coins with lower capitalisation.

    %\begin{itemize}
    %    \item[-] This effect is excacerbated for the median response networks, particularly for the linear-median response subnetwork, where we observe almost 80\% of links pointing to lower capitilisation coins. 
    %\end{itemize} 
\end{itemize}

\vspace{3cm}
\noindent The majority of significant correlations are positive, indicating generally elevated levels of activity for higher capitalisation regions of the networks. The median response networks show relatively high interactions with capitalisation, with many links pointing downstream, and tail-events having larger causal impact for highly capitalised coins. The combination of downstream propensity and negative in-degree correlation for the linear-median network is relatively intuitive, as we would not expect standard return behaviour from low-capitalisation coins to be effecting the largest cryptocurrencies. %For those interested in the detailed set of correlations between the degree types of our sub-networks, see appendix item \autoref{fig:correlation_plot}. The 324 correlation pairs are somewhat difficult to summarise, but we see a general trend of matching degree types positively correlating. This implies that coins have a tendency to be broadly influencing, or influenced, rather than only at specific regions. The comparatively sparser relationships in the non-matching degree types imply that the strength of outgoing and incoming influence is relatively independent per coin. 
For a detailed examination of the correlations between degree types across our subnetworks, refer to Additional Figures item A \autoref{fig:correlation_plot}. While summarizing the 324 correlation pairs is challenging, a general pattern emerges: matching degree types tend to correlate positively. This suggests that coins are more likely to be broadly influential or influenced across the networks, rather than confined to specific regions. In contrast, the relatively sparse correlations between non-matching degree types indicate that the strength of outgoing and incoming influence is largely independent for each coin.

%The strongest observed correlations, those for matching-tails In-Degree distributions 

%Overall, we make the following qualitative interpretations: (1) Median-response information flows "downstream" from significant, large capitalisation influencers. (2) Scale/Quantile-response information flows "Across", with generally elevated activity for higher capitalised coins.  

Overall, we make the following qualitative interpretations: (1) Tail-events in high capitalisation coins tend to have the most outgoing influence. (2). Median-response information has a slight tendency to flow downstream from large capitalisation influences, while scale/quantile-response information flows across.

%\subsection{Analysis of All-Effects Multigraph}
\subsection{All-Effects Multigraph}

%We can also construct a single "all-effects" network as the sum of all subnetwork. The product will be a multigraph, with the possibility of multiple edges between nodes. 

%Figure \ref{fig:multi} displays the CCDF for the this Multigraph, with Table \ref{tb:multi} showing summary statistics. We observe that despite many of the sub-network degree types being correlated to market capitalisation, the multigraph remains largely un-correlated. This highlights the importance of stratefied results, as it shows that the results analysed together obscure the lack of market capitlisation effect. From figure \ref{fig:multi} we observe that despite the correlations somewhat "cancelling out", the relative scaling found in out degree remains somewhat present. 

We can also construct a single all-effects network by summing all subnetworks, resulting in a multigraph where multiple edges between nodes are possible. For comparison against this general causality network we also construct a `baseline' linear causality network by fitting a standard linear-VAR model. For consistency with our existing sub-networks, this standard VAR model is fit against the median target, which is a small departure from the traditional description shown in Section \ref{sec:VAR}. We also compare against a Piecewise VAR network, which is the sum of the three median response networks.

%We compare this general causality to a linear causality network generated by 

%VAR network, which fits a straight linear effect to the median response. 

\begin{figure}[ht!]
    \centering
    \begin{subfigure}[b]{0.32\textwidth}
        \centering
        \includegraphics[width=\textwidth]{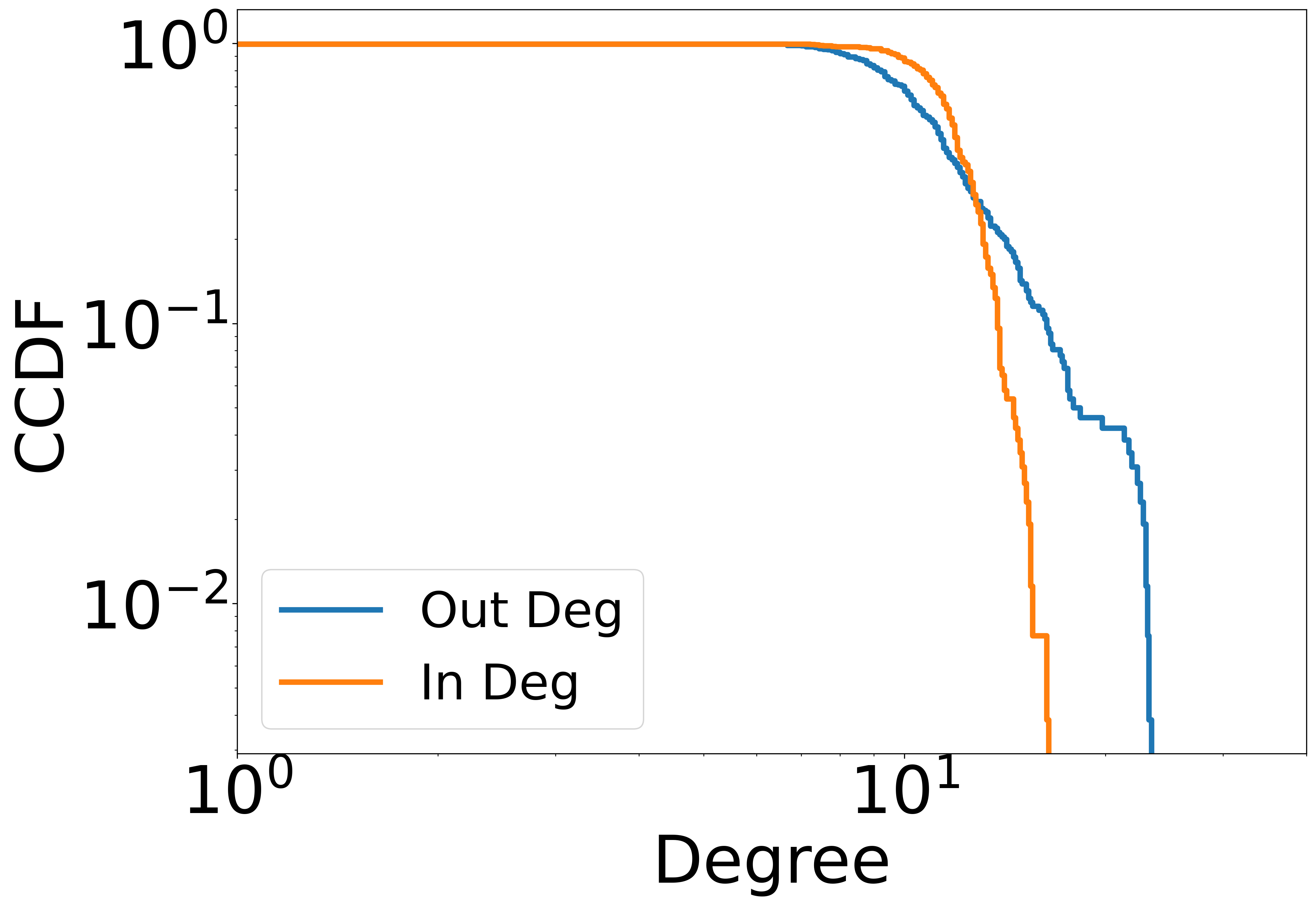} % Replace with your image filename
        \caption{Multigraph}
        \label{fig:1}
    \end{subfigure}
    \hfill
    \begin{subfigure}[b]{0.32\textwidth}
        \centering
        \includegraphics[width=\textwidth]{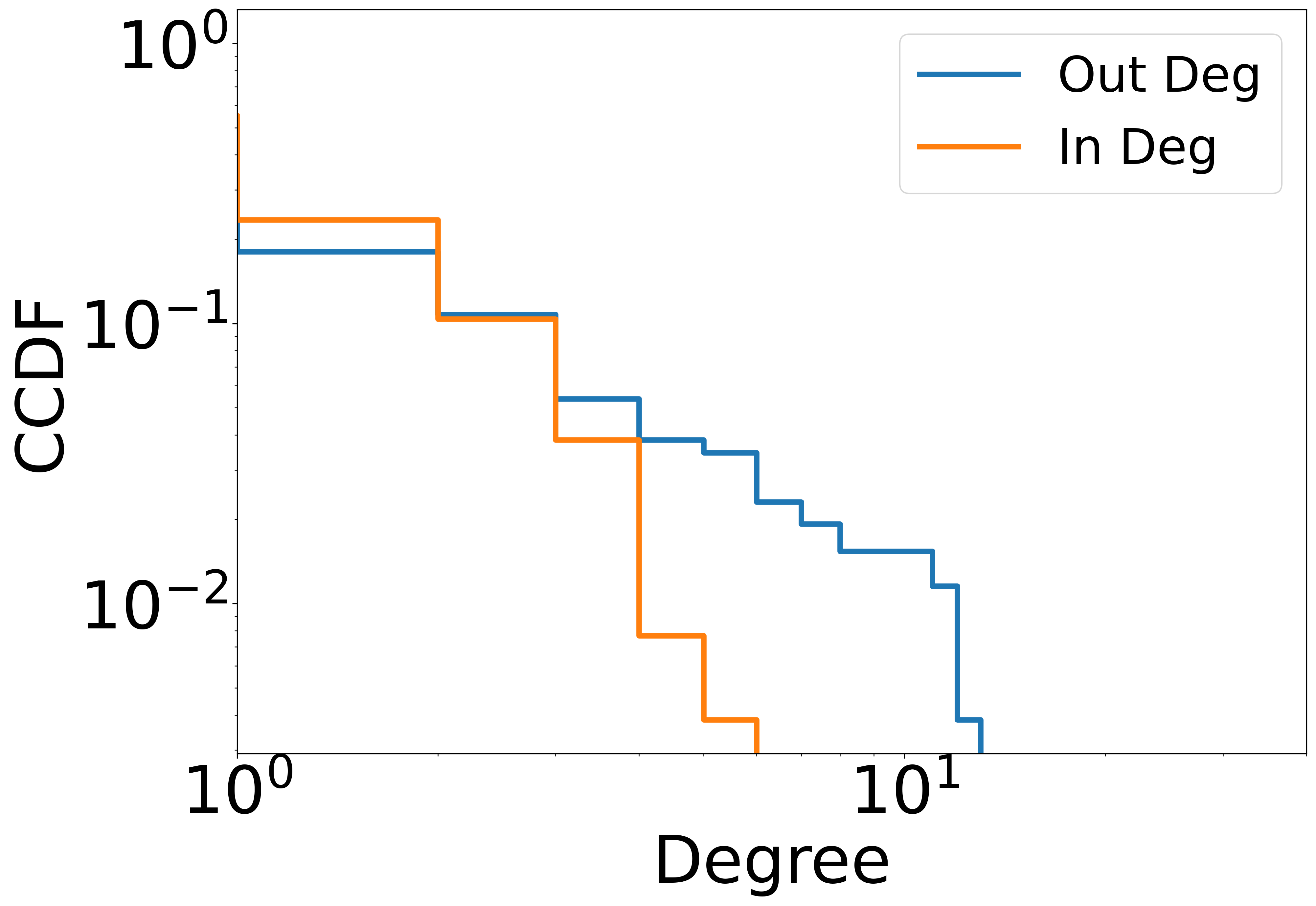} % Replace with your image filename
        \caption{Standard VAR}
        \label{fig:2}
    \end{subfigure}
    \hfill
    \begin{subfigure}[b]{0.32\textwidth}
        \centering
        \includegraphics[width=\textwidth]{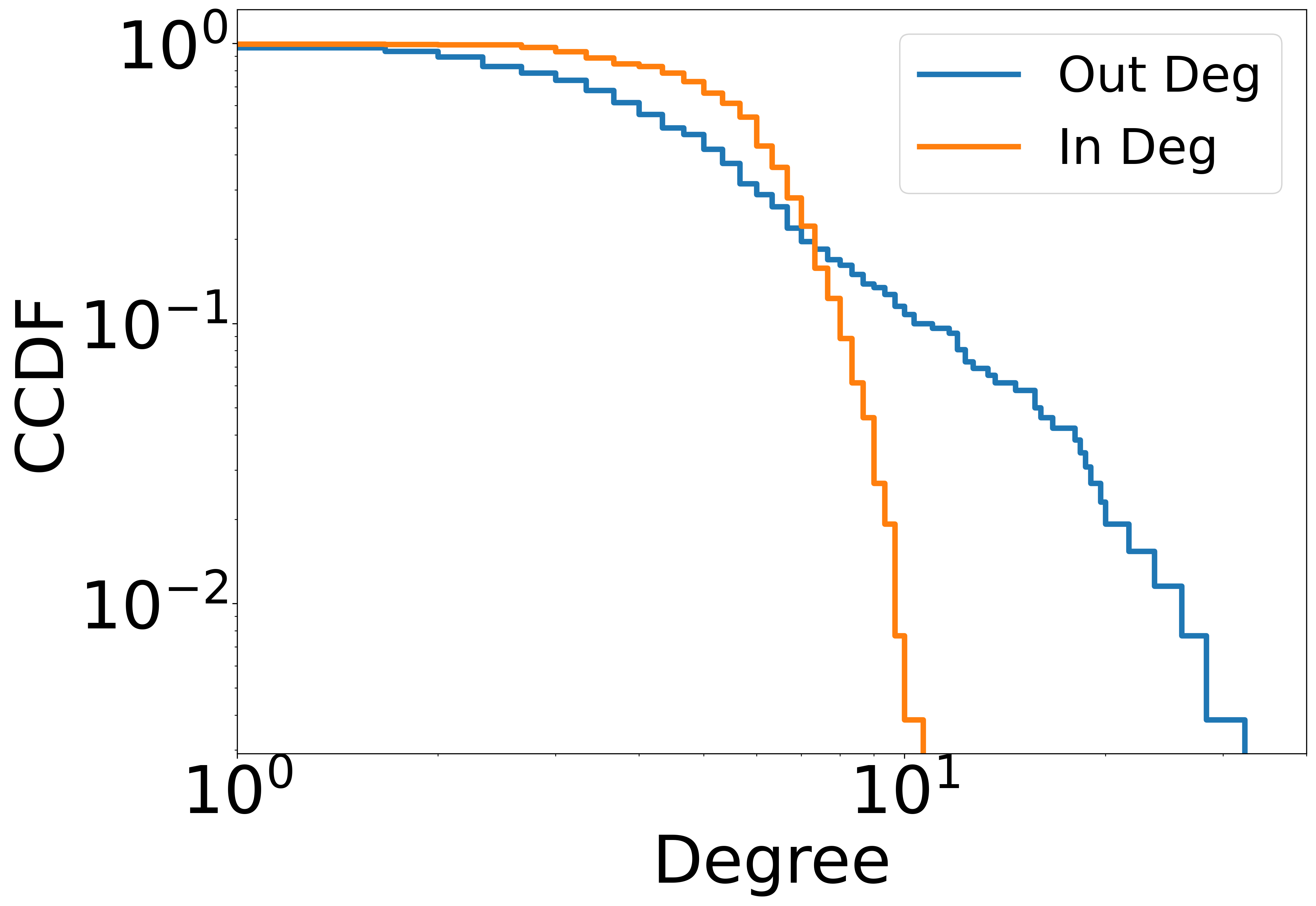} % Replace with your image filename
        \caption{Piecewise VAR}
        \label{fig:3}
    \end{subfigure}
    \caption{Multigraph, Standard VAR and piecewise VAR network node degree CCDFs.}
    \label{fig:strawman_ccdf}
\end{figure}

Figure \ref{fig:strawman_ccdf} showcases the CCDF for these networks, while Table \ref{fig:strawman_table} provides summary statistics (we divide the multigraph degrees by 9, and the piecewise network by 3 to be on an equivalent scale). The most stark difference is the number of identified links. The multigraph, which considers causality across a variety of variable scenarios identifies around 6 times the number of causal links compared to the standard VAR. Even restricting our attention to the median response, we see that splitting the linearity into our piecewise structure leads to 3 times as many identified relationships. Both the multigraph and standard VAR networks lack a statistically significant relationship between out-degree and capitalisation, with the standard model having a significant negative relationship for in-degree, and the piecewise network having a significant relationship for out degree. While the correlation structure for the standard VAR model resembles the linear-median response sub-network we previously identified, it surprisingly contains even fewer links than this sub-network. It would appear that rather than the tail-median effects producing false positives for linear-median effects, they actually obscure/cancel the linear-median effects. A visual example of this could be seen in the Figure 1b, where the significant right tail-median positive effect could obscure the linear-median negative effect.
%Notably, even though many sub-network degree types are correlated with market capitalisation, the multigraph largely remains uncorrelated. 
Comparing the multigraph against the results in Table \ref{tb:3x3} underscores the significance of stratified results, emphasizing that combining the results can mask the absence of a market capitalisation effect. Similarly, the non-stratified CCDF analysis under-emphasises the out-degree scaling we see in the individual sub-networks. 

%These comparisons underscores the significance of stratified results, emphasizing that combining the results can cancel local effects, and mask the presence of market capitalisation effects within different compnents of the variable densities.. 

%An observation from Figure \ref{fig:multi} is that, although the correlations seem to neutralize each other to a degree, the relatively elevated scaling observed for the out-degree is still discernible. 

%This becomes particularly evident when we consider Bitcoin, which has the highest total out-degree (247) among all analysed cryptocurrencies. 

%It is largely unsurprising that across both 

It is unsurprising that within the all-effects multigraph we observe Bitcoin to have the highest total out-degree. This intuitive finding indicates that Bitcoin has statistically significant causal effects at 10.5\% of all possible outgoing links, and 64.6\% of all coins have some causal dependency on Bitcoin. This effect is largely driven by a very high out-degree for the negative-tail to median response subnetwork, where 104 (40\%) coins displayed negative causal effect (100\% negative), indicating that negative outliers for bitcoin tended to be followed by elevated returns for most other assets in the network. %One potential explanation for this phenomena is that significant 

\vspace{-10mm}

\begin{figure}[ht!]
    \centering
    \begin{subfigure}[t!]{0.31\textwidth}
        \centering
        \def\arraystretch{1.25}%
        %\resizebox{\columnwidth}{!}{
        \begin{tabular}[t!]{ l S[table-format=1.4] S[table-format=2.4] l} 
        \hline \textbf{Attribute} & \textbf{mean} &  $\sigma_{\text{attribute}}$ & $\rho$ \\
        \hline
         Out-Deg. & 12.30   & 3.58 & 0.0844 \\
         In-Deg.  & 12.30  & 0.06 & 0.101 \\
         P(Down) & 0.501 & {-} & {-} \\
        \hline
        \end{tabular}
        %}
        \caption{Multigraph}
        \label{tb:1}
    \end{subfigure}
    \hfill
    \begin{subfigure}[t!]{0.32\textwidth}
        \centering
        \def\arraystretch{1.25}%
        %\resizebox{\columnwidth}{!}{
        \begin{tabular}[t]{ l S[table-format=1.4] S[table-format=2.4] l} 
        \hline \textbf{Attribute} & \textbf{mean} &  $\sigma_{\text{attribute}}$ & $\rho$ \\
        \hline
         Out-Deg. & 1.89   & 1.96 & 0.0777 \\
         In-Deg.  & 1.89  & 1.19 & -0.218\smallasttrip\\
         P(Down) & 0.602 & {-} & {-} \\
        \hline
        \end{tabular}
        %}
        \caption{Standard VAR}
        \label{tb:2}
    \end{subfigure}
    \hfill
    \begin{subfigure}[t!]{0.31\textwidth}
        \centering
        \def\arraystretch{1.25}%
        %\resizebox{\columnwidth}{!}{
        \begin{tabular}[t]{ l S[table-format=1.4] S[table-format=2.4] l} 
        \hline \textbf{Attribute} & \textbf{mean} &  $\sigma_{\text{attribute}}$ & $\rho$ \\
        \hline
        Out-Deg. & 6.21   & 5.06 & 0.200 \\
        In-Deg.  & 6.21  & 1.74 & 0.008 \\
        P(Down) & 0.540 & {-} & {-} \\
        \hline
        \end{tabular}
        %}
        \caption{Piecewise VAR}
        \label{tb:3}
    \end{subfigure}
    \caption{Multigraph, Standard VAR and piecewise VAR network summary statistics.}
    \label{fig:strawman_table}
\end{figure}

% \caption{\textcolor{AdBlue}{Empirical network node-specific metrics, showing node degrees, $k^{+-\Delta}$ and the clustering co-efficient}. For each of these attributes, Tables \ref{tbl:mean_network} and \ref{tab:volnets} show the empirical: mean, median, standard deviation $\sigma$, Spearman's rank correlation ($\rho$) with market capitalisation and the statistical significance of independence from capitalization, with significance levels indicated by asterisks($*\rightarrow p\leq0.05, **\rightarrow p\leq0.01, ***\rightarrow p\leq0.001$).}

%\end{document}

%We also observe that higher capitalisation coins have fewer incoming Linear-Median effects. 

%\newpage
\section{Conclusion}
This study introduces a technique for constructing nonlinear distributional causality networks, enabling a direct investigation of tail-tail causal interactions in financial data. By combining piecewise linear embeddings with quantile regressions, we robustly estimate a set of subnetworks representing distinct, locally linear causal effects across key regions of the input and output densities. In an empirical study on 260 cryptocurrencies, our P-QVAR model identifies intricate self-causality structures, with cryptocurrencies displaying an asymmetric blend of mean-reversion, GARCH-like behaviour, and autoregression in the residual skew. We find that large positive events generally exert more causal influence, showcasing heightened mean reversion and skew transference compared to negative events. Cross-causal effects are more symmetrical, with GARCH like behaviour being the most pronounced influence structure. There is comparatively limited cross-causality for the median response target, indicating that most cross-asset coupling occurs in the higher distributional moments. Our findings have implications for both investors and policymakers, as it highlights the relatively independent action of raw returns may not imply isolated behaviour during tail events and extreme market behaviour. This study serves as a foundation for further research into distributional causality networks, prompting a shift in the focus of financial networks from mean-response causality to more risk-inclusive measures.

\newpage
\section*{Additional Figures}
\appendix
\section{ Degree Correlation Matrix}

\begin{figure*}[ht!]
     \centering
         \includegraphics[width=\textwidth]{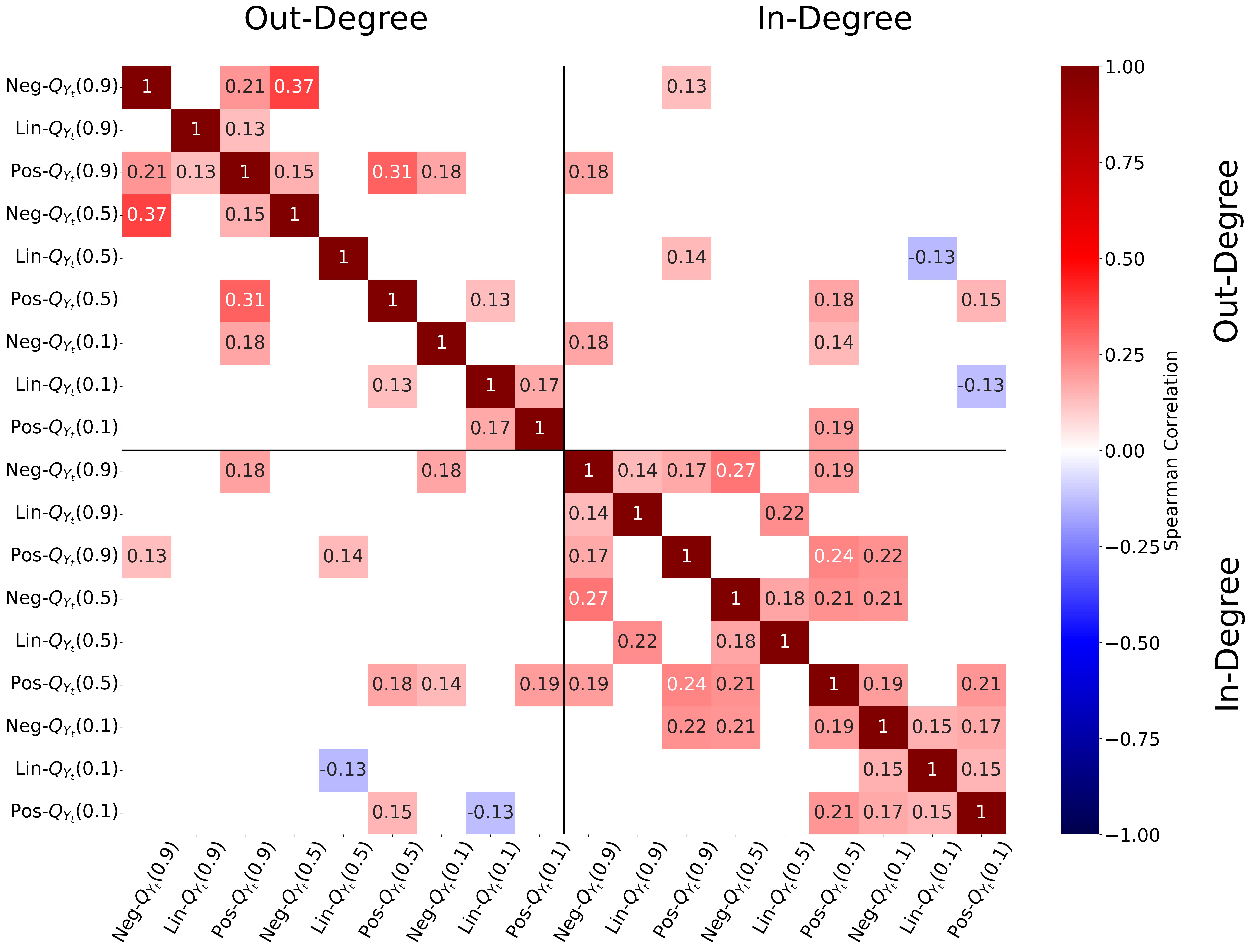}
         %\vspace{-1cm}
         \caption{Correlation heatmap: displaying all statistically significant Spearman correlations between the in-degree and out-degree for each pair of the 9 subnetworks.}
         \label{fig:correlation_plot}
\end{figure*}

\clearpage
%\end{thebibliography}
% ---- OR ------- Uncomment this section to use bibtex
\bibliographystyle{spmpsci} % We choose the "plain" reference style
\bibliography{refs} % Entries are in the refs.bib file
\end{document}

%% file: UL.tex
\def\arraystretch{1.25}%
\resizebox{\columnwidth}{!}{
\begin{tabular}[t]{ l S[table-format=1.4] S[table-format=2.4] l} 
\hline \textbf{Attribute} & \textbf{mean} &  $\sigma_{\text{attribute}}$ & $\rho$ \\
\hline
Out-Deg. & 15.7   & 8.41 &  -0.071\\
In-Deg.  & 15.7  & 4.18 & -0.0435\\
%Clust.  & 0.239  & 0.134 & 0.304 \\
P(Down) & 0.487 & {-} & {-}  \\

\hline
\end{tabular}
}

%% file: UM.tex
\def\arraystretch{1.25}%
\resizebox{\columnwidth}{!}{
\begin{tabular}[t]{ l S[table-format=1.4] S[table-format=2.4] l} 
\hline \textbf{Attribute} & \textbf{mean} &  $\sigma_{\text{attribute}}$ & $\rho$ \\
\hline
Out-Deg. & 12.65   & 4.82 &  0.0223\\
In-Deg.  & 12.65  & 3.84 & -0.0699\\
%Clust.  & 0.239  & 0.134 & 0.304 \\
P(Down) & 0.508 & {-} & {-} \\
\hline
\end{tabular}
}

%% file: UR.tex
\def\arraystretch{1.25}%
\resizebox{\columnwidth}{!}{
\begin{tabular}[t]{ l S[table-format=1.4] S[table-format=2.4] l} 
\hline \textbf{Attribute} & \textbf{mean} &  $\sigma_{\text{attribute}}$ & $\rho$ \\
\hline
Out-Deg. & 16.09   & 7.81 &  0.0364\\
In-Deg.  & 16.09  & 4.45 & 0.217\smallasttrip\\
%Clust.  & 0.239  & 0.134 & 0.304 \\
P(Down) & 0.491 & {-} & {-} \\
\hline
\end{tabular}
}

%% file: ML.tex
\def\arraystretch{1.25}%
\resizebox{\columnwidth}{!}{
\begin{tabular}[t]{ l S[table-format=1.4] S[table-format=2.4] l} 
\hline \textbf{Attribute} & \textbf{mean} &  $\sigma_{\text{attribute}}$ & $\rho$ \\
\hline
Out-Deg. & 7.81   & 11.26 &  0.174\smallastdub  \\
In-Deg.  & 7.81  & 3.12 & 0.00185\\
%Clust.  & 0.239  & 0.134 & 0.304 \\
P(Down) & 0.527 & {-} & {-}  \\
\hline
\end{tabular}
}

%% file: MM.tex
\def\arraystretch{1.25}%
\resizebox{\columnwidth}{!}{
\begin{tabular}[t]{ l S[table-format=1.4] S[table-format=2.4] l} 
\hline \textbf{Attribute} & \textbf{mean} &  $\sigma_{\text{attribute}}$ & $\rho$ \\
\hline
Out-Deg. & 4.36   & 5.36 &  0.0774\\
In-Deg.  & 4.36  & 1.87 & -0.140\smallast\\
%Clust.  & 0.239  & 0.134 & 0.304 \\
P(Down) & 0.544 & {-} & {-}  \\
\hline
\end{tabular}
}

%% file: MR.tex
\def\arraystretch{1.25}%
\resizebox{\columnwidth}{!}{
\begin{tabular}[t]{ l S[table-format=1.4] S[table-format=2.4] l} 
\hline \textbf{Attribute} & \textbf{mean} &  $\sigma_{\text{attribute}}$ & $\rho$ \\
\hline
Out-Deg. & 6.46   & 7.42 &  0.144\smallast\\
In-Deg.  & 6.46  & 2.68 & 0.0978\\
%Clust.  & 0.239  & 0.134 & 0.304 \\
P(Down) & 0.553 & {-} & {-} \\
\hline
\end{tabular}
}

%% file: LL.tex
\def\arraystretch{1.25}%
\resizebox{\columnwidth}{!}{
\begin{tabular}[t]{ l S[table-format=1.4] S[table-format=2.4] l} 
\hline \textbf{Attribute} & \textbf{mean} &  $\sigma_{\text{attribute}}$ & $\rho$ \\
\hline
Out-Deg. & 16.54   & 9.99 & 0.170\smallastdub  \\
  In-Deg.  & 16.54  & 4.478 & 0.205\smallasttrip  \\
%  Clust.  & 0.239  & 0.134 & 0.304 \\
P(Down) & 0.50 & {-} & {-}  \\
\hline
\end{tabular}
}

%% file: LM.tex
\def\arraystretch{1.25}%
\resizebox{\columnwidth}{!}{
\begin{tabular}[t]{ l S[table-format=1.4] S[table-format=2.4] l} 
\hline \textbf{Attribute} & \textbf{mean} &  $\sigma_{\text{attribute}}$ & $\rho$ \\
\hline
Out-Deg. & 12.25   & 6.63 &  -0.0206\\
In-Deg.  & 12.25  & 3.56 & 0.102\\
%Clust.  & 0.239  & 0.134 & 0.304 \\
P(Down) & 0.493 & {-} & {-}  \\
\hline
\end{tabular}
}

%% file: LR.tex
\def\arraystretch{1.25}%
\resizebox{\columnwidth}{!}{
\begin{tabular}[t]{ l S[table-format=1.4] S[table-format=2.4] l} 
\hline \textbf{Attribute} & \textbf{mean} &  $\sigma_{\text{attribute}}$ & $\rho$ \\
\hline
Out-Deg. & 15.89   & 9.69 &  -0.128\smallast\\
In-Deg.  & 15.89  & 4.24 & -0.0099\\
%Clust.  & 0.239  & 0.134 & 0.304 \\
P(Down) & 0.486 & {-} & {-}  \\
\hline
\end{tabular}
}

%% file: main.bbl
\begin{thebibliography}{10}
\providecommand{\url}[1]{{#1}}
\providecommand{\urlprefix}{URL }
\expandafter\ifx\csname urlstyle\endcsname\relax
  \providecommand{\doi}[1]{DOI~\discretionary{}{}{}#1}\else
  \providecommand{\doi}{DOI~\discretionary{}{}{}\begingroup \urlstyle{rm}\Url}\fi

\bibitem{ahelegebey}
Ahelegbey, D.F., Billio, M., Casarin, R.: Bayesian graphical models for structural vector autoregressive processes.
\newblock Journal of Applied Econometrics \textbf{31}(2), 357--386 (2016)

\bibitem{Ahelegbey2021NetworkBE}
Ahelegbey, D.F., Cerchiello, P., Scaramozzino, R.: Network based evidence of the financial impact of {Covid}-19 pandemic.
\newblock International Review of Financial Analysis \textbf{81}, 102,101 -- 102,101 (2021)

\bibitem{ALYAHYAEE2018228}
Al-Yahyaee, K.H., Mensi, W., Yoon, S.M.: Efficiency, multifractality, and the long-memory property of the bitcoin market: A comparative analysis with stock, currency, and gold markets.
\newblock Finance Research Letters \textbf{27}, 228--234 (2018).
\newblock \doi{https://doi.org/10.1016/j.frl.2018.03.017}.
\newblock \urlprefix\url{https://www.sciencedirect.com/science/article/pii/S1544612318300242}

\bibitem{GARCHN}
Almansour, B., Alshater, M., Almansour, A.: Performance of arch and garch models in forecasting cryptocurrency market volatility.
\newblock Industrial Engineering \& Management Systems \textbf{20}, 130--139 (2021).
\newblock \doi{10.7232/iems.2021.20.2.130}

\bibitem{structural_entropy}
Almog, A., Shmueli, E.: Structural entropy: Monitoring correlation-based networks over time with application to financial markets.
\newblock Scientific Reports \textbf{9}, 10,832 (2019)

\bibitem{amirzadeh_nazari_thiruvady_ee}
Amirzadeh, R., Nazari, A., Thiruvady, D., Ee, M.S.: Modelling determinants of cryptocurrency prices: A bayesian network approach  (2023).
\newblock \urlprefix\url{https://ssrn.com/abstract=4403923}.
\newblock Working paper, available at SSRN

\bibitem{connecting_emotions}
Aste, T.: Cryptocurrency market structure: connecting emotions and economics.
\newblock Digital Finance \textbf{1} (2019)

\bibitem{covid_indicator}
Azimli, A.: The impact of covid-19 on the degree of dependence and structure of risk-return relationship: A quantile regression approach.
\newblock Finance Research Letters \textbf{36}, 101,648 (2020).
\newblock \doi{https://doi.org/10.1016/j.frl.2020.101648}.
\newblock \urlprefix\url{https://www.sciencedirect.com/science/article/pii/S1544612320304815}

\bibitem{AZQUETAGAVALDON2020122574}
Azqueta-Gavaldón, A.: Causal inference between cryptocurrency narratives and prices: Evidence from a complex dynamic ecosystem.
\newblock Physica A: Statistical Mechanics and its Applications \textbf{537}, 122,574 (2020)

\bibitem{BALCILAR201674}
Balcilar, M., Gupta, R., Pierdzioch, C.: Does uncertainty move the gold price? new evidence from a nonparametric causality-in-quantiles test.
\newblock Resources Policy \textbf{49}, 74--80 (2016).
\newblock \doi{https://doi.org/10.1016/j.resourpol.2016.04.004}.
\newblock \urlprefix\url{https://www.sciencedirect.com/science/article/pii/S0301420716300563}

\bibitem{BAUR2013786}
Baur, D.G.: The structure and degree of dependence: A quantile regression approach.
\newblock Journal of Banking \& Finance \textbf{37}(3), 786--798 (2013).
\newblock \doi{https://doi.org/10.1016/j.jbankfin.2012.10.015}.
\newblock \urlprefix\url{https://www.sciencedirect.com/science/article/pii/S0378426612003263}

\bibitem{billio2011}
Billio, M., Lo, A., Sherman, M., Pelizzon, L.: Econometric measures of connectedness and systemic risk in the finance and insurance sectors.
\newblock Journal of Financial Economics \textbf{104} (2011)

\bibitem{stat_analysis_of_network}
Boginski, V., Butenko, S., Pardalos, P.: Statistical analysis of financial networks.
\newblock Computational Statistics \& Data Analysis \textbf{48}, 431--443 (2005)

\bibitem{CHEIKH2020101293}
Cheikh, N.B., Zaied, Y.B., Chevallier, J.: Asymmetric volatility in cryptocurrency markets: New evidence from smooth transition garch models.
\newblock Finance Research Letters \textbf{35}, 101,293 (2020).
\newblock \doi{https://doi.org/10.1016/j.frl.2019.09.008}.
\newblock \urlprefix\url{https://www.sciencedirect.com/science/article/pii/S154461231930162X}

\bibitem{chu}
Chu, J., Chan, S., Nadarajah, S., Osterrieder, J.: Garch modelling of cryptocurrencies.
\newblock Journal of Risk and Financial Management \textbf{10}, 17 (2017).
\newblock \doi{10.3390/jrfm10040017}

\bibitem{CHUANG20091351}
Chuang, C.C., Kuan, C.M., Lin, H.Y.: Causality in quantiles and dynamic stock return–volume relations.
\newblock Journal of Banking \& Finance \textbf{33}(7), 1351--1360 (2009).
\newblock \doi{https://doi.org/10.1016/j.jbankfin.2009.02.013}.
\newblock \urlprefix\url{https://www.sciencedirect.com/science/article/pii/S0378426609000235}

\bibitem{cornell2023vector}
Cornell, C., Mitchell, L., Roughan, M.: Vector autoregression in cryptocurrency markets: Unraveling complex causal networks (2023).
\newblock ArXiv preprint arXiv:2308.15769

\bibitem{rank2}
Cornell, C., Mitchell, L., Roughan, M.: Rank is all you need: development and analysis of robust causal networks.
\newblock Applied Network Science \textbf{9}(1), 39 (2024).
\newblock \doi{10.1007/s41109-024-00648-w}.
\newblock \urlprefix\url{https://doi.org/10.1007/s41109-024-00648-w}

\bibitem{conf}
Cornell, C., Mitchell, L., Roughan, M.: Rank is all you need: Robust estimation of complex causal networks.
\newblock In: Complex Networks {\&} Their Applications XII, pp. 468--482. Springer Nature Switzerland, Cham (2024)

\bibitem{Elsayed2020CausalityAD}
Elsayed, A.H., Gozgor, G., Lau, C.K.M.: Causality and dynamic spillovers among cryptocurrencies and currency markets.
\newblock International Journal of Finance \& Economics  (2020)

\bibitem{assymetric2}
Engle, R.F., Manganelli, S.: Caviar: Conditional autoregressive value at risk by regression quantiles.
\newblock Journal of Business \& Economic Statistics \textbf{22}(4), 367--381 (2004).
\newblock \doi{10.1198/073500104000000370}.
\newblock \urlprefix\url{https://doi.org/10.1198/073500104000000370}

\bibitem{FAKHFEKH2020101075}
Fakhfekh, M., Jeribi, A.: Volatility dynamics of crypto-currencies’ returns: Evidence from asymmetric and long memory garch models.
\newblock Research in International Business and Finance \textbf{51}, 101,075 (2020).
\newblock \doi{https://doi.org/10.1016/j.ribaf.2019.101075}.
\newblock \urlprefix\url{https://www.sciencedirect.com/science/article/pii/S027553191930056X}

\bibitem{greene2012econometric}
Greene, W.H.: Econometric Analysis, 7th, international edition edn.
\newblock Pearson, Boston (2012)

\bibitem{GUO2018251}
Guo, P., Zhu, H., You, W.: Asymmetric dependence between economic policy uncertainty and stock market returns in g7 and bric: A quantile regression approach.
\newblock Finance Research Letters \textbf{25}, 251--258 (2018).
\newblock \doi{https://doi.org/10.1016/j.frl.2017.11.001}.
\newblock \urlprefix\url{https://www.sciencedirect.com/science/article/pii/S1544612317306864}

\bibitem{hseather}
Hall, P., Sheather, S.J.: On the distribution of a studentized quantile.
\newblock Journal of the Royal Statistical Society. Series B (Methodological) \textbf{50}(3), 381--391 (1988).
\newblock \urlprefix\url{http://www.jstor.org/stable/2345702}

\bibitem{JENA2019615}
Jena, S.K., Tiwari, A.K., Hammoudeh, S., Roubaud, D.: Distributional predictability between commodity spot and futures: Evidence from nonparametric causality-in-quantiles tests.
\newblock Energy Economics \textbf{78}, 615--628 (2019).
\newblock \doi{https://doi.org/10.1016/j.eneco.2018.11.013}.
\newblock \urlprefix\url{https://www.sciencedirect.com/science/article/pii/S0140988318304535}

\bibitem{long_range_VAR}
Johansen, S.: A representation theory for a class of vector autoregressive models for fractional processes.
\newblock Econometric Theory \textbf{24}, 651--676 (2008)

\bibitem{assymetric1}
Karim, M.M., Kawsar, N.H., Ariff, M., Masih, M.: Does implied volatility (or fear index) affect islamic stock returns and conventional stock returns differently? wavelet-based granger-causality, asymmetric quantile regression and nardl approaches.
\newblock Journal of International Financial Markets, Institutions and Money \textbf{77}, 101,532 (2022).
\newblock \doi{https://doi.org/10.1016/j.intfin.2022.101532}.
\newblock \urlprefix\url{https://www.sciencedirect.com/science/article/pii/S1042443122000233}

\bibitem{partial_corr}
Kenett, D., Tumminello, M., Madi, A., Gershgoren, G., Mantegna, R., Ben-Jacob, E.: Dominating clasp of the financial sector revealed by partial correlation analysis of the stock market.
\newblock PloS one \textbf{5}, e15,032 (2010)

\bibitem{tvals0}
Koenker, R., Bassett, G.: Regression quantiles.
\newblock Econometrica \textbf{46}(1), 33--50 (1978).
\newblock \urlprefix\url{http://www.jstor.org/stable/1913643}

\bibitem{tvals1}
Koenker, R., Bassett, G.: Robust tests for heteroscedasticity based on regression quantiles.
\newblock Econometrica \textbf{50}(1), 43--61 (1982).
\newblock \urlprefix\url{http://www.jstor.org/stable/1912528}

\bibitem{QAR}
Koenker, R., Xiao, Z.: Quantile autoregression.
\newblock Journal of the American Statistical Association \textbf{101}(475), 980--990 (2006).
\newblock \doi{10.1198/016214506000000672}.
\newblock \urlprefix\url{https://doi.org/10.1198/016214506000000672}

\bibitem{luet}
Luetkepohl, H.: The New Introduction to Multiple Time Series Analysis (2005)

\bibitem{crypto1}
Milunovich, G.: Cryptocurrencies, mainstream asset classes and risk factors - a study of connectedness (2018)

\bibitem{clustering_and_info}
Onnela, J.P., Kaski, K., Kertész, J.: Clustering and information in correlation based financial networks.
\newblock The European Physical Journal B - Condensed Matter \textbf{38} (2003)

\bibitem{rogers1993calculation}
Rogers, W.: Calculation of quantile regression standard errors.
\newblock Stata technical bulletin no. 13, Stata Corporation, College Station, TX (1993)

\bibitem{seabold2010statsmodels}
Seabold, S., Perktold, J.: Statsmodels: Econometric and statistical modeling with python.
\newblock In: 9th Python in Science Conference (2010)

\bibitem{sims_1980}
Sims, C.: Macroeconomics and reality.
\newblock Econometrica \textbf{48}(1), 1–48 (1980)

\bibitem{Combiningsocial_and_financial}
Souza, T., Aste, T.: Predicting future stock market structure by combining social and financial network information.
\newblock Physica A: Statistical Mechanics and its Applications \textbf{535}, 122,343 (2019)

\bibitem{NAR}
Zhu, X., Pan, R., Li, G., Liu, Y., Wang, H.: {Network vector autoregression}.
\newblock The Annals of Statistics \textbf{45}(3), 1096 -- 1123 (2017).
\newblock \doi{10.1214/16-AOS1476}.
\newblock \urlprefix\url{https://doi.org/10.1214/16-AOS1476}

\bibitem{ZHU2019345}
Zhu, X., Wang, W., Wang, H., Härdle, W.K.: Network quantile autoregression.
\newblock Journal of Econometrics \textbf{212}(1), 345--358 (2019).
\newblock \doi{https://doi.org/10.1016/j.jeconom.2019.04.034}.
\newblock \urlprefix\url{https://www.sciencedirect.com/science/article/pii/S0304407619300892}.
\newblock Big Data in Dynamic Predictive Econometric Modeling

\end{thebibliography}
